\documentclass[11pt]{article}
\usepackage{amssymb}
\usepackage{amsfonts}
\usepackage{graphicx}
\usepackage{amsmath}
\usepackage[pdfa]{hyperref}

\makeatletter \@addtoreset{equation}{section} \makeatother

\newtheorem{theorem}{Theorem}[section]
\newtheorem{lemma}{Lemma}[section]

\newtheorem{definition}{Definition}[section]
\newtheorem{proposition}{Proposition}[section]

\newcommand{\mdet}{\mathrm{det}}
\newcommand{\intd}{\displaystyle\int}
\newcommand{\Tr}{\mathrm{Tr}\,}

\begin{document}


\title{Transfer matrix approach for the real symmetric 1D random band matrices}
\author{
 Tatyana Shcherbina
 \thanks{Department of Mathematics,  University of Wisconsin-Madison, Madison, WI, USA, e-mail: tshcherbyna@wisc.edu. }
\thanks{Department of Mathematics, Princeton University, Princeton, NJ, USA. Supported in part by NSF grant DMS-1700009.}}

\date{}
\maketitle
\begin{abstract}
This paper adapts the recently developed rigorous application of the supersymmetric transfer matrix approach for the 1d band matrices 
to the case of the orthogonal symmetry.
We consider $N\times N$ block band matrices consisting of $W\times W$ random Gaussian blocks (parametrized by  $j,k \in\Lambda=[1,n]\cap \mathbb{Z}$, $N=nW$) with a
fixed entry's variance $J_{jk}=W^{-1}(\delta_{j,k}+\beta\Delta_{j,k})$ in each block.  Considering the limit $W, n\to\infty$, we prove that the behavior  
of the second correlation function of characteristic polynomials of such matrices in the bulk of the spectrum exhibit a crossover near the threshold $W\sim \sqrt{N}$.

\end{abstract}

\section{Introduction}
Starting from the works of Erd$\ddot{\hbox{o}}$s,
Yau, Schlein with coauthors (see \cite{EYY:10} and reference therein)  and Tao and Vu (see, e.g., \cite{TaoVu:11}),
 the significant progress in understanding the universal behavior of many random graph and random matrix
models were achieved. However  for the random  matrices with spacial structure our understanding is much more limited.

One of the most important  models with a spatial structure are random band matrices (RBM), which are interpolating  model
 between mean-field type Wigner
matrices (Hermitian or real symmetric matrices with i.i.d.  random entries) and random Schr$\ddot{\hbox{o}}$dinger operators, 
which have only a random diagonal potential in addition to the deterministic Laplacian on a box in $\mathbb{Z}^d$. 

The main long standing problem in the field is to prove a  fundamental  physical conjecture formulated in late 80th (see \cite{Ca-Co:90}, \cite{FM:91}). 
The conjecture states that  the eigenvectors of $N\times N$ RBM are completely delocalized and the local  spectral  statistics  
governed  by  the Wigner-Dyson statistics  
for large bandwidth $W$ (i.e. the local behavior is the same as for Wigner matrices), and by Poisson statistics for a small $W$ (with exponentially localized eigenvectors). The transition is conjectured to be sharp and for RBM 
in  one  spatial  dimension  occurs around  the  critical  value $W=\sqrt{N}$. This  is  the  analogue  of  the  celebrated  Anderson  
metal-insulator  transition  for random Schr$\ddot{\hbox{o}}$dinger operators.

The conjecture on the crossover in RBM with $W\sim\sqrt N$ is supported by physical derivation due to Fyodorov and Mirlin (see \cite{FM:91}) based on supersymmetric formalism, and also by the so-called Thouless scaling. On the mathematical level of rigour,
localization of eigenvectors in the bulk of the spectrum was first shown for $W\ll N^{1/8}$ \cite{S:09}, and then the bound was improved to $N^{1/7}$  \cite{wegb:16} .
On the other side, 
by a development of the Erd\H{o}s-Schlein-Yau approach to Wigner matrices (see \cite{EYY:10}),  there were obtained some results where the weaker form of delocalization was 
proved for  $W\gg N^{6/7}$ in \cite{EK:11} , $W\gg N^{4/5}$ in \cite{Yau:12}, $W\gg N^{7/9}$ in \cite{HM:18}.
The combination of this approach with the new ideas based on quantum unique ergodicity gave first GUE/GOE gap distributions for RBM with $W\sim N$ \cite{BEYY:16}, 
and then were developed in \cite{BYY:18} --  \cite{BYY2:18}, \cite{BYY3:18} to obtain bulk universality and complete delocalization in the  range  $W\gg N^{3/4}$ (see review \cite{B:rev} for the details).

There is a completely different approach  which allows to work with random operators with non-trivial spatial
structures based on supersymmetry techniques (SUSY). It is widely used in the physics literature (see e.g. reviews
\cite{Ef},\cite{M:00}) but  its rigorous mathematical application is usually quite difficult
and it requires to incorporate various analytic and statistical mechanics techniques. 
However for the Hermitian RBM of a certain type it was successfuly done both for correlation functions of characteristic polynomials
and for usual correlation functions. More precisely, combining SUSY with a delicate steepest descent method
and transfer matrix techniques, we were able to perform a complete study of the local regime
of characteristic polynomials for Hermitian Gaussian 1d RBM  (see \cite{TSh:14} for the regime $W\gg \sqrt{N}$, \cite{SS:ChP_loc} for the regime $W\ll \sqrt{N}$,
and \cite{TSh:ChP_crit} for the regime $W\sim \sqrt{N}$), and also to obtain the first rigorous universality result for the second order correlation function for
the whole delocalized region $W\gg \sqrt{N}$ (see \cite{SS:Un}). 

There are much less rigorous application of SUSY techniques for the case of real symmetric matrices, since
 the SUSY integral representations are more complicated for the
case of orthogonal symmetry. However,  the techniques of \cite{TSh:14} were successfully adapted in \cite{TSh:ChP_sym} 
to the study of characteristic polynomials for real symmetric Gaussian 1d RBM in the delocalized regime $W\gg \sqrt{N}$.
In this paper we want to perform the complete study of characteristic polynomials for real symmetric Gaussian 1d RBM
adapting the SUSY transfer matrix techniques of \cite{SS:ChP_loc}, \cite{TSh:ChP_crit} to the case of orthogonal symmetry.
This  is an important step
towards the proof of the universality of the usual correlation functions for the case of real symmetric
1d RBM, as well as for the general development of rigour application of SUSY approach for the real symmetric case.

The model we are going to consider is different from the model of 1d RBM considered in \cite{TSh:14} -- \cite{SS:ChP_loc}, \cite{TSh:ChP_crit} and
in \cite{TSh:ChP_sym}, but coincides with the model considered in \cite{SS:Un}. Namely, we consider real symmetric block band matrices, i.e. real symmetric matrices $H_N$, $N=n W$ with elements $H_{jk,\alpha\gamma}$,
where $j,k \in 1,\ldots, n$ (they parametrize the lattice sites) and $\alpha, \beta= 1,\ldots, W$ (they
parametrize the orbitals on each site). The entries $H_{jk,\alpha\gamma}$ are random
Gaussian variables with mean zero such that
\begin{equation}\label{H}
\langle H_{j_1k_1,\alpha_1\beta_1}H_{j_2k_2,\alpha_2\beta_2}\rangle=\delta_{j_1k_2}\delta_{j_2k_1}
\delta_{\alpha_1\beta_2}\delta_{\beta_1\alpha_2} J_{j_1k_1}.
\end{equation}
Here $J_{jk}\ge 0$ are matrix elements of the positive-definite symmetric $n\times n$ matrix $J$, such that
\[
\sum\limits_{j=1}^nJ_{jk}=1/W.
\]
The probability law of $H_N$ can be written in the form
\begin{equation}\label{pr_l}
P_N(d H_N)=\exp\Big\{-\dfrac{1}{4}\sum\limits_{j,k\in\Lambda}\sum\limits_{\alpha,\gamma=1}^W
\dfrac{H_{jk,\alpha\gamma}^2}{J_{jk}}\Big\}dH_N,
\end{equation}
where
\[
dH_N=\prod\limits_{j<k}\prod\limits_{\alpha\gamma}\dfrac{dH_{jk,\alpha\gamma}}{\sqrt{2\pi J_{jk}}}
\prod\limits_{j}\prod\limits_{\alpha<\gamma}\dfrac{dH_{jj,\alpha\gamma}}{\sqrt{2\pi J_{jj}}}
\prod\limits_{j}\prod\limits_{\alpha}\dfrac{dH_{jj,\alpha\alpha}}{\sqrt{4\pi J_{jj}}}.
\]
Such models were first introduced and studied by Wegner (see \cite{S-W:80}, \cite{We:79}) (and sometimes also called
Wegner's orbital models).

We restrict ourselves to the case
\begin{equation}\label{J}
J=1/W+\beta\Delta^{(0)}/W, \quad \beta<1/4,
\end{equation}
where $W\gg 1$ and $\Delta^{(0)}$ is the discrete Laplacian on $[1,n]\cap \mathbb{Z}$ with Neumann boundary conditions.
Clearly, this model  is one of the possible realizations of the Gaussian random band matrices with the band width $2W+1$
(note that the model can be defined similarly in any dimensions $d>1$   taking  $j,k\in [1,n]^d\cap \mathbb{Z}^d$
in (\ref{H})).

For 1D RBM it was shown
in \cite{BMP:91, MPK:92} that $\mathcal{N}_{N}$ converges weakly, as $N,W\to\infty$, to a non-random measure
$\mathcal{\mathcal{N}}$, which is called the limiting NCM of the ensemble. The measure $\mathcal{N}$ is absolutely continuous
and its density $\rho$ is given by the well-known Wigner semicircle law :
\begin{equation}\label{rho_sc}
\rho(E)=\dfrac{1}{2\pi}\sqrt{4-E^2},\quad E\in[-2,2].
\end{equation}
In this paper we consider  the
correlation functions (or the mixed moments) of characteristic polynomials, which can be defined as
\begin{equation}\label{F}
F_{2k}(\Lambda)=\intd \prod\limits_{s=1}^{2k}\mdet(\lambda_s-H_N)P_n(d\,H_N),
\end{equation}
where $P_n(d\,H_N)$ is defined in (\ref{pr_l}),
and $\Lambda=\hbox{diag}\,\{\lambda_1,\ldots,\lambda_{2k}\}$ are real or complex parameters
that may depend on $N$.  As in the Hermitian case, correlation functions of characteristic polynomials of real symmetric 1d RBM are expected to exhibit a crossover 
near the threshold $W\sim \sqrt{N}$: it is expected that they the same local behavior 
as for GOE for $W\gg \sqrt{N}$, and the different behavior for $W\ll \sqrt{N}$.

The asymptotic local behavior in the bulk of the spectrum of the $2k$-point mixed moment
for GOE is well-known. It was proved for $k=1$ by Br$\acute{\hbox{e}}$zin and Hikami
\cite{Br-Hi:01} (based on SUSY approach), and for general $k$ by Borodin and Strahov \cite{BorSt:06} (with a different
techniques) that
\begin{equation*}
F_{2k}\left(\Lambda_0+\hat{\xi}/n\rho(E)\right)
=C_{N,k}\dfrac{\hbox{Pf}\,
\big\{DS(\pi(\xi_i-\xi_j))
\big\}_{i,j=1}^{2k}}{\triangle(\xi_1,\ldots,\xi_{2k})}(1+o(1)),
\end{equation*}
where $C_{N,k}$ is some multiplicative constant depending on $N$, $k$,
\begin{equation}\label{dS}
DS(x)=-\dfrac{3}{x}\dfrac{d}{dx}\dfrac{\sin x}{x}=3\Big(\dfrac{\sin x}{x^3}-
\dfrac{\cos x}{x^2}\Big),
\end{equation}
$\triangle(\xi_1,\ldots,\xi_k)$ is the
Vandermonde determinant of $\xi_1,\dots, \xi_k$, and $$\hat{\xi}=\hbox{diag}\,\{\xi_1,\ldots,\xi_{2k}\},\quad
\Lambda_0=E\cdot I.$$
In particular, for $k=1$ we have
\begin{equation*}
F_{2}\left(\Lambda_0+\hat{\xi}/n\rho(E)\right)
=C_N\Big(\dfrac{\sin (\pi (\xi_1-\xi_2))}{\pi^3 (\xi_1-\xi_2)^3}-
\dfrac{\cos (\pi (\xi_1-\xi_2))}{\pi^2 (\xi_1-\xi_2)^2}\Big)(1+o(1)),
\end{equation*}
The last formula was proved also  for real symmetric Wigner and general sample covariance matrices
(see \cite{Kos:09}).

Set
\begin{equation}\label{lambda}
\lambda_1=E+\dfrac{\xi}{2N\rho(E)},\quad \lambda_2=E-\dfrac{\xi}{2N\rho(E)},
\end{equation}
where $E\in (-2,2)$, $\rho$ is defined in (\ref{rho_sc}), and
$\xi$ is a real parameter varying in any compact
set $K\subset \mathbb{R}$, and define
\begin{equation}\label{D_2}
D_2=F_2^{1/2}(E,E).
\end{equation}

 The main result of the paper is the following theorem :
\begin{theorem}\label{thm:1}
For the real symmetric 1d block random band matrices $H_N$, $N=nW$ of (\ref{H}) -- (\ref{J}) we have 
\begin{equation*}
\lim\limits_{n\to\infty} 
\bar F_{2}\Big(E+\dfrac{\xi}{2N\rho(E)},E-\dfrac{\xi}{2N\rho(E)}\Big)=
\left\{
\begin{array}{cc}
DS(\pi\xi),& W\gg n\gg 1;\\
(e^{-C^*\Delta- i\xi\pi\hat\nu}\cdot 1,1),& n=C_*W\\
1,&  1\ll W\le n/\log^2 n,
\end{array}
\right.
\end{equation*}
where $DS(x)$ is defined in (\ref{dS}), $C^*=C_*/(2\pi\rho(E))^2$ with $\rho(E)$ of (\ref{rho_sc}), and $\varepsilon$ is any sufficiently small positive number. In this formula  
$\Delta$ is a Laplace-Bertrami  operator  on $\mathring{Sp}(2)=Sp(2)/Sp(1)\times Sp(1)$, and
$(\cdot,\cdot)$ is an inner product on $L_2[\mathring{Sp}(2), d\mu]$, where $d\mu$ is the Haar measure on 
$\mathring{Sp}(2)$. $\hat\nu$ is an operator of multiplication by
\begin{align}\label{nu}
 \nu(Q)=1-2(|Q_{12}|^2+|Q_{14}|^2)
\end{align}
on $\mathring{Sp}(2)$. Notice that the since $N=nW$, the transition happens at $W\sim \sqrt{N}$.
\end{theorem}

\subsection{Notation}
We denote by $C$, $C_1$, etc. various $W$ and $N$-independent quantities below, which
can be different in different formulas. Integrals
without limits denote the integration (or the multiple integration) over the whole
real axis, or over the Grassmann variables.

Moreover,
\begin{itemize}

\item $W$ is a size of the block, and $n$ is the number of blocks in a row, so $N=nW$ is the size of the matrix $H$ of (\ref{H});

\item $\mathbf{E}\big\{\ldots\big\}$ is an expectation with respect to the measure (\ref{pr_l});

\item
$\hfill
 a_{\pm}=\dfrac{iE\pm\sqrt{4-E^2}}{2}=e^{\pm i\alpha_0};\hfill
$\label{a_pm}

\item
 $
\hfill \sigma=\left(\begin{array}{cc}
0&1\\
-1&0
\end{array}\right),\quad
\sigma'=\left(\begin{array}{cc}
0&1\\
1& 0
\end{array}\right); \hfill
$\label{s-s'}

\item
$\hfill
D_0=\left(\begin{array}{cc}
a_+&0\\
0& a_-
\end{array}\right),\quad
\Lambda=\left(\begin{array}{cc}
\lambda_1&0\\
0& \lambda_2
\end{array}\right),\quad
\hat{\xi}=\left(\begin{array}{cc}
\xi&0\\
0& -\xi
\end{array}\right),\quad
L=\left(\begin{array}{cc}
1&0\\
0& -1
\end{array}\right); \hfill
$
\item
$ \hfill
D_{0,4}=\left(\begin{array}{cc}
D_0&0\\
0&D_0
\end{array}\right),\quad
\hat{\xi}_4=\left(\begin{array}{cc}
\hat{\xi}&0\\
0& \hat{\xi}
\end{array}\right),\quad
L_4=\left(\begin{array}{cc}
L&0\\
0& L
\end{array}\right); \hfill
$
\label{xi_4}

\item $\hfill \Lambda_0=E\cdot I_2, \quad \Lambda_{0,4}=E\cdot I_4;\hfill $\label{l_0}

\item $U(n)$ is a group of $n\times n$ unitary matrices; unitary symplectic group $Sp(n)$ is a group of $2n\times 2n$ unitary matrices $Q$ which admit the relation
\[
Q\, \left(\begin{array}{cc}
0&I_n\\
-I_n&0
\end{array}\right)\, Q^{t}=\left(\begin{array}{cc}
0&I_n\\
-I_n&0
\end{array}\right).
\]

\item $\mathring{U}(2)=U(2)/\big(U(1)\times U(1)\big)$, \quad $\mathring{Sp}(2)=Sp(2)/\big(Sp(1)\times Sp(1)\big);$

\item $\hfill \mathbb{T}=\{z\in\mathbb{C}: |z|=1\},\quad \omega_A=\{z\in\mathbb{C}: |z|=1+A/n\};\hfill $ 

\item $d\mu$ is the Haar measure on $\mathring{U}(2)$, $d\mu$ is the Haar measure on $\mathring{Sp}(2)$;

\item
$ \hfill
 c_\pm=1+a_\pm^{-2};\quad t_*=(2\pi\rho(E))^2\hfill
$\label{c_pm}

\item We denote by $\bar a$ the vector $(a_1,a_2)$;
%
%
%
%
\end{itemize}

\section{Integral representation}
The main aim of this section is to derive the following proposition
\begin{proposition}\label{p:int_repr}
The second correlation function  (\ref{F}) of the characteristic polynomials for 1d real symmetric Gaussian block band
matrices (\ref{H}) -- (\ref{J}) can be represented as follows:
\begin{align}\label{int_repr}
&F_2\Big(\Lambda_0+\dfrac{\hat{\xi}}{2N\rho(E)}\Big)=C_{n,W}
\int\exp\Big\{\frac{\beta W}{4}
\sum\limits_{j=2}^n\Tr\,(F_j-F_{j-1})^2\Big\}\\ \notag
&\times\exp\Big\{\frac{W}{4}\sum\limits_{j=1}^n\Big(\Tr F_j^2-2i\Tr F_j\big(\Lambda_{0,4}+\dfrac{\hat{\xi}_4}{2N\rho(E)}\big)\Big)\Big\}
\prod\limits_{j=1}^n
\big(\mdet F_j\big)^{-W/2}\prod\limits_{j=1}^ndF_j,
\end{align}
where $\Lambda_{0,4}$ and $\hat{\xi}_4$ are defined in Notation, $N=nW$, $C_{nW}$ is some constant depending on $W$ and $n$ but not on $\xi$, and
$F_j\in Sp(2)$ are unitary symplectic $4\times 4$ matrices.
\end{proposition}
\textit{Proof.}
Introduce the following Grassmann fields:
\begin{align*}
\Psi_l=\{\psi_{jl}^t\}^t_{j=1,..,n}, &\quad \psi_{j l}=(\psi_{j l 1}, \psi_{j l 2},\ldots,
\psi_{j l W})^t,\quad l=1,2.
\end{align*}
Using (\ref{G_Gr}) (see Appendix A) we obtain
\begin{equation*}
\begin{array}{c}
F_2(\lambda_1,\lambda_2)
=\mathbf{E}\Big\{\displaystyle\int \exp\{-\Psi_1^+(\lambda_1-H_N)\Psi_1
-\Psi_2^+(\lambda_2-H_N)\Psi_2\} d\Psi\Big\}\\
=\displaystyle\int  d\Psi\,\, \exp\Big\{-\lambda_1\Psi_1^+\Psi_1
-\lambda_2\Psi_2^+\Psi_2\Big\}\\
\times \mathbf{E}\Big\{\exp\Big\{\sum\limits_{j< k}\sum\limits_{\alpha, \gamma}
H_{jk,\alpha\gamma}(\eta_{jk,\alpha\gamma}+\eta_{kj,\gamma\alpha})+\sum\limits_{j}\sum\limits_{\alpha\le \gamma}
H_{jk,\alpha\gamma}(\eta_{jk,\alpha\gamma}+\eta_{kj,\gamma\alpha})\Big\}\Big\},
\end{array}
\end{equation*}
where 
\begin{align*}
&
d\Psi=\prod\limits_{j=1}^n\prod\limits_{\alpha=1}^W\prod\limits_{l=1}^2d\overline{\psi}_{jl\alpha}
d\psi_{j l\alpha},\\
&\eta_{jk,\alpha\gamma}=\overline{\psi}_{j 1\alpha}\psi_{k 1\gamma}+
\overline{\psi}_{j 2\alpha}\psi_{k 2\gamma},\quad \hbox{if}\,\,j\ne k\,\,\hbox{or}\,\,\alpha\ne\gamma;\\
&\eta_{jj,\alpha\alpha}=(\overline{\psi}_{j 1\alpha}\psi_{j 1\alpha}+
\overline{\psi}_{j 2\alpha}\psi_{j 2\alpha})/2.
\end{align*}
Averaging over (\ref{pr_l}), we get
\begin{equation*}
\begin{array}{c}
F_2(\lambda_1,\lambda_2)=\displaystyle\int  d\Psi\,\, \exp\{-\lambda_1\Psi_1^+\Psi_1
-\lambda_2\Psi_2^+\Psi_2\} \\
\times\exp\Big\{\frac{1}{2}\sum\limits_{j<k, \alpha,\gamma} J_{jk}\,\,
(\eta_{jk,\alpha\gamma}+\eta_{kj,\gamma\alpha})^2+\frac{1}{2}\sum\limits_{j, \alpha<\gamma}J_{jj}\,(\eta_{jj,\alpha\gamma}+\eta_{jj,\gamma\alpha})^2+
\sum\limits_{j, \alpha} J_{jj}\,\,
\eta_{jj,\alpha\alpha}^2\Big\}.
\end{array}
\end{equation*}
It is easy to see that
\begin{align*}
\frac{1}{2}\sum\limits_{\alpha,\gamma}(\eta_{jk,\alpha\gamma}+\eta_{kj,\gamma\alpha})^2=&-(\psi_{j1}^+\bar\psi_{j2})(\psi_{k1}^t\psi_{k2})
-(\psi_{k1}^+\bar\psi_{k2})(\psi_{j1}^t\psi_{j2})-(\psi_{j1}^+\psi_{j1})(\psi_{k1}^+\psi_{k1})\\&-(\psi_{j2}^+\psi_{j2})(\psi_{k2}^+\psi_{k2})-(\psi_{j1}^+\psi_{j2})( \psi_{k2}^+\psi_{k1})-(\psi_{k1}^+\psi_{k2})(\psi_{j2}^+\psi_{j1})\\
=&-\frac{1}{2}\Tr \tilde F_j\tilde F_k,\\
\sum\limits_{l=1,2}\lambda_l\Psi_l^+\Psi_l=&\dfrac{1}{2}\sum\limits_{j=1}^n\Tr \tilde F_j\Lambda_4
\end{align*}
where
\begin{equation*}
\tilde F_j=\left(\begin{array}{cccc}
\psi_{j1}^+\psi_{j1}&\psi_{j1}^+\psi_{j2}&0&\psi_{j1}^+\bar\psi_{j2}\\
\psi_{j2}^+\psi_{j1}&\psi_{j2}^+\psi_{j2}&\psi_{j2}^+\bar\psi_{j1}&0\\
0&\psi_{j1}^t\psi_{j2}&\psi_{j1}^+\psi_{j1}&\psi_{j2}^+\psi_{j1}\\
\psi_{j2}^t\psi_{j1}&0&\psi_{j1}^+\psi_{j2}&\psi_{j2}^+\psi_{j2}
\end{array}\right),\quad
\Lambda_4=\left(\begin{array}{cccc}
\lambda_1&0&0&0\\
0&\lambda_2&0&0\\
0&0&\lambda_1&0\\
0&0&0&\lambda_2
\end{array}\right).
\end{equation*}
Applying the superbosonization formula (see Proposition \ref{p:supboz}, Appendix A) we obtain
\begin{align}\notag
F_2(\lambda_1,\lambda_2)&=C'_{nW}\displaystyle\int \exp\Big\{-\dfrac{1}{4}\sum\limits_{j,k=1}^nJ_{jk}\Tr F_jF_k-\dfrac{1}{2}\sum\limits_{j,k=1}^n\Tr F_j\Lambda_4\Big\}\prod\limits_{j=1}^n(\mdet F_j )^{-W/2}\prod\limits_{j=1}d\,F_j,
\end{align}
where $\{F_j\}_{j=1}^n$ are unitary symplectic $4\times 4$ matrices form $Sp(2)$, and $C'_{nW}$ is some constant depending on $W$ and $n$ but not on $\lambda_1$, $\lambda_2$. Shifting $F_j\to iWF_j$ and plugging in (\ref{lambda}), we get Proposition \ref{p:int_repr}.

$\Box$

\section{Representation in the operator form}

To study (\ref{int_repr}), we are going to apply the transfer matrix approach.

Namely, introduce
\begin{align}\label{F_cal}
\mathcal{F} (X)&=\exp\Big\{W\Big(
\frac{1}{8}\, \Tr X^2-\dfrac{iE}{4}\Tr X-\frac{1}{4}\,\Tr \log
X-C_+\Big)\Big\},\\ \notag
\mathcal{F}_\xi (X)&=\mathcal{F} (X)\cdot \mathcal{F}_{n,\xi}(X),\quad  \mathcal{F}_{n,\xi}(X):=\exp\Big\{-
\frac{i}{8n\rho(E)}\, \Tr X\hat\xi_4\Big\}
\end{align}
where 
\begin{align*}
C_+=\frac{a_+^2}{2}-iEa_+-\log a_+
\end{align*}
is chosen in such a way that $|\mathcal{F} (X)|=1$ in the saddle-points (see (\ref{st_points_1}) later).

Let also $K, K_\xi: Sp(2)\to Sp(2)$ be the operators with the kernels 
\begin{align} \label{K}
K(X,Y)&=\dfrac{W^3}{2\pi^3}\,\mathcal{F}(X)\,\exp\Big\{\frac{\beta W}{4}\Tr
(X-Y)^2\Big\}\,\mathcal{F}(Y);\\
\label{K_xi}
K_{\xi}(X,Y)&=\dfrac{W^3}{2\pi^3}\,\mathcal{F}_\xi(X)\,\exp\Big\{\frac{\beta W}{4}\Tr
(X-Y)^2\Big\}\,\mathcal{F}_\xi(Y).
\end{align}
Then Proposition \ref{p:int_repr} can be reformulated as
\begin{align}\label{F_rep}
F_2\Big(E+\dfrac{\xi}{2N\rho(E)},E-\dfrac{\xi}{2N\rho(E)}\Big)=\tilde C_{n,W} ( K_\xi^{n-1}\mathcal{F}_{\xi},\bar{\mathcal{F}}_{\xi}),
\end{align}
where $(\cdot,\cdot)$ is a standard inner product in $Sp(2)$ with respect to the Haar measure $d\mu$, and $\tilde C_{nW}$ is some constant depending on $W$ and $n$ but not on $\xi$.

For arbitrary compact operator $M$  denote by $\lambda_j(M)$ the $j$th (by its modulo) eigenvalue
of $M$,
so that $|\lambda_0(M)|\ge|\lambda_1(M)|\ge\dots$. 

 Since $K_\xi$ is a compact operator, one can rewrite 
\begin{align*}\notag
( K_\xi^{n-1}\mathcal{F}_{\xi},\bar{\mathcal{F}}_{\xi})=\sum_{j=0}^\infty\lambda_j^{n-1}(K_\xi)c_j,\quad with\quad
c_j=(\mathcal{F}_{\xi},\psi_j)(\bar{\mathcal{F}}_{\xi},\tilde\psi_j),
\end{align*}
where $\{\psi_j\}$ are  eigenvectors corresponding to $\{\lambda_j(K_\xi)\}$, and $\{\tilde \psi_j\}$ are the eigenvectors of $K^*_\xi$. 
Similar equality is true if we replace $K_\xi$ and $\mathcal{F}_{\xi}$ by $K$ and $\mathcal{F}$.
Hence, to study (\ref{int_repr}),
it suffices to study the eigenvalues and eigenvectors of $K_\xi$, $K$.

\section{Sketch of the proof of Theorem \ref{thm:1}}

As was mentioned above,  we are interested in the analysis of the spectral properties of  $K_\xi$ of (\ref{K_xi}) (see (\ref{F_rep})). 
It appears that
 it is  simpler  to work with the resolvent 
analog of (\ref{F_rep})
\begin{align}\label{res_rep}
(K_\xi^{n-1}f, g)=-\frac{1}{2\pi i}\oint_{\mathcal{L}}z^{n-1}(G_\xi(z)f, g)dz,\quad 
G_\xi(z)=(K_\xi-z)^{-1},
\end{align}
where $\mathcal{L}$ is any closed contour which enclosed all eigenvalues of $K_\xi$. 

The idea of the proof is very close to \cite{SS:ChP_loc} -- \cite{TSh:ChP_crit}. To outline it, we start with the following definition
\begin{definition}\label{def:1}
 We  say that the operator $\mathcal{A}_{n,W}$  is equivalent to  $\mathcal{B}_{n,W}$  ($\mathcal{A}_{n,W}\sim\mathcal{B}_{n,W}$) on some contour
 $\mathcal{L}$ if
\[ \int_{\mathcal{L}}z^{n-1}((\mathcal{A}_{n,W}-z)^{-1}f,\bar g)dz= \int_{\mathcal{L}}z^{n-1}((\mathcal{B}_{n,W}-z)^{-1}f,\bar g) dz \, (1+o(1)),\quad n,W\to \infty,\]
with some particular functions $f,g$ depending of the problem.
\end{definition}
The aim  is to find  some operator equivalent to $ K_\xi$  whose spectral analysis is more accessible.
Now we are going to discuss how this was done on the ideological level.
The specific choice of the contour $\mathcal{L}$  and functions $f$, $g$ for each step will be discussed in details in Section 6.

It is easy to check that the stationary  points of the function $\mathcal{F}$ of  (\ref{F_cal})  are
\begin{align}\label{st_points_1}
X_+&=a_+\cdot I_4,\quad X_-=a_-\cdot I_4;\\
X_\pm(Q)&=\,QD_{0,4}Q^*,\quad Q\in \mathring{Sp}(2) \notag
\end{align}
where $a_\pm$, $D_{0,4}$ are defined in Notation. Notice also that the value of $|\mathcal{F}|$ at points (\ref{st_points_1}) is $1$.

The first step in the proof of Theorem \ref{thm:1}  is to apply the saddle-point approximation. Roughly speaking, we show that if we introduce the projection $Pr_{s}$  onto  the $W^{-1/2}\log W$-neighbourhoods of the saddle points $X_+$, $X_-$ and the saddle "surface" $X_\pm$, then in the sense of Definition \ref{def:1} 
\begin{align*}
K_{\xi}\sim Pr_{s}K_{\xi}Pr_{s}=:K_{s,\xi}.
\end{align*}
Moreover, we can show that only the neighborhood of the saddle "surface" $X_\pm$ gives the main contribution to the integral.
The proof is based on a study of a quadratic approximation of a function $\mathcal{F}$ of (\ref{F_cal}).
Let us also emphasize, that for the block band matrices (\ref{H}) -- (\ref{J}) this step is much simpler than for the model considered in 
\cite{SS:ChP_loc} -- \cite{TSh:ChP_crit} due to the large coefficient $W$ in the exponent of $\mathcal{F}$.
This analysis will be performed in details in Section 5.

To study the operator $K_{s,\xi}$  near the saddle ``surface" $X_\pm$ we  use the  "polar coordinates". Namely,
the matrices from $Sp(2)$ have two eigenvalues $a_j,b_j\in \mathbb{T}=\{z: |z|=1\}$ of the multiplicity two and
can be considered as quaternion $2\times 2$ matrices.
In this language $F_j$ are quaternion unitary  matrices, and so they can be diagonalized
by the quaternion unitary $2\times 2$ matrices from $\mathring{Sp}(2)$ (see , e.g., \cite{Me:91}, Chapter 2.4).

Change the variables to $F_j=Q_j^*A_{j,4} Q_j$, where 
$A_{j,4}=\hbox{diag}\,\{a_{1j},a_{2j},a_{1j},a_{2j}\}$, $a_{1j},a_{2j}\in \mathbb{T}$, and $Q_j\in \mathring{Sp}(2)$.
Then $d F_j$ of (\ref{int_repr}) becomes (see, e.g., \cite{Me:91} )
$$\dfrac{\pi^2}{12}(a_{1j}-a_{2j})^4\,d\bar a_j\,
d\mu(Q_j),$$ where 
\[
d\bar a_j=\dfrac{da_{1j}}{2\pi i}\,\dfrac{da_{2j}}{2\pi i},
\]
and $d\mu(Q_j)$ is the normalized to unity Haar measure on the symplectic group $\mathring{Sp}(2)$.
Thus we get
\begin{align*}
&( K_\xi^{n-1}\mathcal{F}_{\xi},\bar{\mathcal{F}}_{\xi})=\dfrac{\pi^{2n}}{12^n}\int (a_{11}-a_{21})^2 F_\xi(a_{11},a_{21},Q_1)(a_{1n}-a_{2n})^2  F_\xi(a_{1n},a_{2n},Q_n)\\
&\times\prod\limits_{j=1}^{n-1}\Big( 
(a_{1j}-a_{2j})^2(a_{1,j+1}-a_{2,j+1})^2K_\xi(F_j,F_{j+1})\Big) \prod\limits_{j=1}^{n}d\bar a_j\,
d\mu(Q_j). \notag 
\end{align*}
Introduce
\begin{align}\label{t}
t= (a_1-a_2)(a_1'-a_2').
\end{align} 
Then we obtain
\begin{align}\label{F_rep1}
F_2\Big(E+\dfrac{\xi}{2N\rho(E)},E-\dfrac{\xi}{2N\rho(E)}\Big)=\tilde C_{n,W}' ( K_\xi^{n-1}f,\bar f),
\end{align}
where now $(\cdot,\cdot)$ is a standard inner product in $L_2[\mathbb{T}^2]\times L_2[\mathring{Sp}(2),d\mu(Q)]$, and $\tilde C_{nW}'$ is some constant depending on $W$ and $n$ but not on $\xi$.
Here
\begin{align}\label{f,g}
f(a_1,a_2,Q)=(a_{1}-a_{2})^2 F_\xi(a_{1},a_{2},Q),
\end{align}
and $K_{\xi}=F_{n,\xi}KF_{n,\xi}$ is  an integral operator in $L_2[\mathbb{T}^2]\times L_2[\mathring{Sp}(2),d\mu(Q)]$  defined by the kernel
\begin{align}\label{rep_2}
&K_\xi(X,Y)=F_{n,\xi}(a_1,a_2,Q)K(a_1,a_2,Q; a_1', a_2',Q')F_{n,\xi}(a_1',a_2',Q'),
\end{align}
where
\begin{align} \notag
&K(a_1,a_2,Q; a_1', a_2',Q')=A_a(\bar a, \bar a')K_*(t,Q_1,Q_2);\\ \label{K(U)}
&K_*(t,Q,Q'):=\dfrac{\beta^2W^2t^2}{6}\cdot \exp\{-t\beta WS\big(Q(Q')^*\big)\},\quad S(Q)=|Q_{12}|^2+|Q_{14}|^2;\\
&F_{n,\xi}(a,b,Q)=\exp\{-i\xi\pi \cdot \nu(a-b,Q)/n\};\notag\\
&\nu(p,Q)=\frac{p}{4\pi\rho(E)}\,\Tr QL_4Q^*L_4=\frac{p}{2\pi\rho(E)} (1-2 S(Q))
\label{nu,s(Q)}\end{align}
with $t$ of (\ref{t}). $K_*$ here is a contribution of the symplectic group $\mathring{Sp}(2)$ into operator $K$, and $\exp\{-i\xi\pi \cdot \nu(x,Q)/n\}$ comes from 
the $1/n$-order perturbation $\mathcal{F}_{n,\xi}$ of $\mathcal{F}$ appearing in $\mathcal{F}_\xi$ (see (\ref{F_cal})).
Operator $A_a$ is a contribution of eigenvalues $a_1,a_2$ and it has the form
\begin{align}\label{A_a}
&A_a(\bar a; \bar a')=A(a_1,a_1') A(a_2,a_2'),\\\notag
&A(a,a')= \Big(\dfrac{W}{2\pi}\Big)^{1/2}e^{- W\Lambda(a,a')};\\
&\Lambda(x,y)=\frac{\beta }{2}(x-y)^2-\frac{1}{2}\varphi_0(x)-\frac{1}{2}\varphi_0(y)+\Re\varphi_0(a_+);\notag\\
&\varphi_0(x)={x^2}/{2}-ixE-\log x.\label{phi_0} 
\end{align}
Observe that the operator $K_{*}(t, Q,Q')$ with some $t>0$ is  self-adjoint  and  its kernel depends only on $S\big(Q(Q')^*\big)$. Thus by the
standard representation theory arguments (see e.g. \cite{Helg}, \cite{Vil:68}),
its eigenfunctions are the the same as for Laplace-Bertrami operator on $Sp(2)$. More precisely:
  \begin{proposition}\label{p:K(U)}
Consider any self-adjoint integral operator $M$ in $L_2[\mathring{Sp}(2), d\mu (Q)]$. If  its kernel $M(Q,Q')$ depends only on $Q(Q')^*$,
then its eigenvectors coincide with eigenvectors of Laplace-Bertrami operator on $\mathring{Sp}(2)$. Moreover,
if the subspace 
\begin{align*}
L_2[S, d\mu (Q)]\subset L_2[\mathring{Sp}(2), d\mu (Q)]
\end{align*}
 of the functions depending on $S(Q)$ (see (\ref{K(U)})) only is invariant under $M$, then 
it can be diagonalized by the eigenfunctions 
\begin{equation}\label{phi_j}
\phi_{j}(Q)=(-1)^jP_{2j}(\sqrt{S(Q)}),
\end{equation}
 where $P_{2j}(x)$ are orthogonal with
respect to the weight $(1-x^2)x^3$on $[0,1]$ polynomials of degree $2j$, $\phi_0(x)=1$ (polynomials $P_{2j}$ can be written as 
$P_{2j}(x)=c_jF_{hg}(-j,j+3,2;1-x^2)$, where $F_{hg}$ is a hypergeometric function, and $c_j$ is a normalization constant , see \cite{Helg}, Ch. 5).  In addition, the following  holds
\begin{align}\label{3_diag}
(2x^2-1)P_{2j}(x)=\dfrac{j+3}{2j+3}P_{2j+2}(x)+\dfrac{j}{2j+3}P_{2j-2}(x),
\end{align}
so the operator $\hat\nu$ of multiplication on $\nu(x,Q)$ of (\ref{nu,s(Q)}) is three diagonal in basis (\ref{phi_j}), and
\begin{align}\label{nu1,1}
(\hat\nu \cdot \phi_0,\phi_0)=0.
\end{align}
If $M(Q_1,Q_2)=K_*(t,Q_1,Q_2)$ of (\ref{K(U)}), then the corresponding
 eigenvalues $\{\lambda_{j}(t)\}_{j=0}^\infty$, if  $t>d>0$, where $d$ is some absolute positive constant, have the form
 \begin{align}\label{l_j}
&\lambda_{j}(t)=1-\frac{(j+1)(j+2)}{W t}+O((j^2/W t)^2) +O(e^{-tW}).
\end{align}
\end{proposition}
The proof of the proposition can be found in Appendix B.

Notice that since 
 $\mathcal F(Q)$, $\mathcal{F}_\xi(Q)$  are the functions of  $S(Q)$ only, and hence according to Proposition \ref{p:K(U)}
in what follows we can consider  restrictions of $K_\xi$, $K$, and $K_*$ of  (\ref{K(U)}) to $L_2[S, d\mu(Q)]$ (to simplify notations we will denote these
restrictions by the same letters).

In addition, it follows from Proposition \ref{p:K(U)} that if we introduce the following basis in $L_2[\mathbb{R}^2]\times L_2[S, d\mu(Q)]$ 
\begin{align*}
&\Psi_{\bar k,j}(\bar a,Q)=\Psi_{\bar k}(\bar a)\phi_{j}(Q), \\
&\Psi_{\bar k}(\bar a)=\psi_{k_1}(a_1)\psi_{k_2}(a_2),
\notag\end{align*}
where $\bar k=(k_1,k_2)$, and $\{\psi_{k}(x)\}_{k=0}^\infty$ is a certain basis in $L_2[\mathbb{R}]$, then the matrix of
$K$ of (\ref{K(U)}) in this basis has a ``block diagonal  structure", which means that
\begin{align}\label{block}
&(K\Psi_{\bar k',j},\Psi_{\bar k, j_1})=0,\quad j\not=j_1\\
&(K\Psi_{\bar k',j},\Psi_{\bar k, j})=(K_j\Psi_{\bar k'},\Psi_{\bar k})\notag\\
=&\int \lambda_{j}(t)A_a(\bar a,\bar a')\psi_{k_1}(a_1)\psi_{k_2}(a_2)
\psi_{k_1'}(a_1')\psi_{k_2'}(a_2')\dfrac{da_1da_2da_1'da_2'}{(2\pi i)^4}.
\notag\end{align}
The next step in the proof of Theorem {\ref{thm:1}} is to show that we can restrict the number of $\phi_j$ to 
\begin{equation}\label{l}
l=\max\{1, [\log n\cdot \sqrt{W}/\sqrt{n}]\}.
\end{equation}
$l$ is chosen in such a way that $l^2n/W\gg \log n$.
 More precisely, we are going to show that
in the sense of Definition \ref{def:1}
\begin{equation*}
K_{s,\xi}\sim \mathcal{P}_{l}K_{s,\xi}\mathcal{P}_{l}=:K_{s,l,\xi},
\end{equation*}
where $\mathcal{P}_{l}$ is the projection on the linear span of $\{\Psi_{\bar k,j}(\bar{a},Q)\}_{j\le l-1}$.

For the further resolvent analysis we want to integrate out $\bar a$ to change  $t$ in the definition of $K_*$ and $a_1-a_2$, $a_1'-a_2'$ in the definition of $F_{n,\xi}$
(see (\ref{t}), (\ref{rep_2}) -- (\ref{K(U)})) by their saddle-point values $t_*=(a_+-a_-)^2=4\pi^2\rho(E)^2$ and $a_+-a_-=2\pi\rho(E)$ correspondingly.
We are going to show that only top eigenvalue of $A$ gives the contribution.
More precisely we want to show that 
in the sense of Definition \ref{def:1}
\begin{equation}\label{tens_pr}
\lambda_0(K_{s,l})^{-1}K_{s,l,\xi}\sim \hat K_{*\xi,l}
\end{equation}
where
\begin{align}\label{K_*xi}
&K_{*\xi,l}=(\lambda_0(K_{*0}))^{-1}P_l\,K_{*\xi}\,P_l,\\
\notag &K_{*\xi}(Q_1,Q_2)=\dfrac{W^2t_*^2\beta^2}{6}\cdot e^{-\beta t_*W S(Q_1Q_2^*)}\cdot e^{-i\xi\pi (\nu(2\pi\rho(E),Q_1)+\nu(2\pi\rho(E),Q_2))/n}
\end{align}
and $P_l$ is the projection on  $\{\phi_j(Q)\}_{j\le l-1}$. Here $K_{* 0}$ is $K_{*\xi}$ with $\xi=0$.

 Now (\ref{tens_pr}), (\ref{res_rep}) and Definition \ref{def:1} give
\begin{multline*}
F_{2}\Big(E+\dfrac{\xi}{2N\rho(E)},E-\dfrac{\xi}{2N\rho(E)}\Big)=C_{n,W}\Big(\hat K_{*\xi,l}^{n-1}f_\xi,\bar f_\xi\Big)(1+o(1))\\
=C_{n,W}\lambda_0(K_{s,l})^{n-1}f_0^2(\hat K_{*\xi,l}^{n-1}1,1)(1+o(1)),
\end{multline*}
where $f_0=(f,\Psi_{\bar 0})$, and we used that $f_\xi$ asymptotically can be replaced by $ f\otimes 1$, where $f$ does not depend on $\xi$ and $Q_j$.
Similarly
\[
D_2=C_{n,W}\Big(\hat K_{*0,l}^{n-1}f_\xi,\bar f_\xi\Big)(1+o(1))\\
=C_{n,W}\lambda_0(K_{s,l})^nf_0^2(\hat K_{*0,l}^{n-1}1,1)(1+o(1)).
\] 
According to Proposition \ref{p:K(U)}, $\phi_0(Q)=1$ is an eigenvector of $\hat K_{*0}$ of (\ref{K_*xi}) with $\xi=0$ and the corresponding eigenvalue is  $1$,  thus
\begin{equation*}
(\hat K_{*0,l}^{n-1}1,1)= 1.
\end{equation*}
Hence 
\begin{equation}\label{ratio}
\bar F_{2}\Big(E+\dfrac{\xi}{2N\rho(E)},E-\dfrac{\xi}{2N\rho(E)}\Big)=(\hat K_{*\xi,l}^{n-1}1,1)(1+o(1)).
\end{equation}
Recall that according to Proposition \ref{p:K(U)} the eigenvectors of $\hat K_{*0,l}$ are (\ref{phi_j}) and the corresponding eigenvalues are (see (\ref{l_j}))
\begin{align}\label{la_j}
\lambda_j:=\lambda_j(t_*)=1-j(j+3)/t_*W+O((j(j+3)/W)^2),\quad j=0,1\dots,l-1.
\end{align}
Moreover, it follows from (\ref{rep_2}) -- (\ref{K(U)}) that
\begin{equation}\label{K_exp}
\hat K_{*\xi,l}=\hat K_{*0,l}-n^{-1}\pi i\xi\hat\nu_l+o(n^{-1}),\quad \hat\nu_l=P_l\hat\nu P_l,
\end{equation} 
where $\hat\nu$ is the operator of multiplication by (\ref{nu}), and $o(1/n)$ means some operator whose norm is $o(1/n)$. Thus  the eigenvalues of  $\hat K_{*\xi,l}$ are in the $n^{-1}$-neighbourhood of
$\lambda_j$. 

In the localized regime $W^{-1}\gg  n^{-1}$ we have $l=1$, 
thus only $\lambda_0(\hat K_{*\xi})$ gives the contribution to (\ref{ratio}).
Since (see Proposition \ref{p:K(U)})
\[(\hat \nu \,1,1)=0,\]
we get
\[
\lambda_0(\hat K_{*\xi})=1+o(n^{-1}), 
\]
and so the limit of (\ref{ratio}) is $1$ (see the end of Section 6 for more details).

In the regime of delocalization  all eigenvalues of $K_{*\xi,l}$ give contribution to
(\ref{ratio}), but $K_{*0,l}^{n-1}\to I$ (roughly speaking, this means
that the second term in the r.h.s. of (\ref{la_j}) does not give a contribution).  Hence we have
\[ \hat K_{*\xi,l}\approx 1-n^{-1}i\xi\pi\hat \nu_l \Rightarrow (K_{*\xi}^{n-1}\,1,1)\to (e^{-i\xi\pi\hat\nu}1,1)=DS(\pi\xi)
\]
with $DS(\pi\xi)$ of (\ref{dS}) (see (\ref{Int_del}) and the end of Section 6 for more details).

In the critical regime $W^{-2}=C_*n^{-1}$ all eigenvalues of $\hat K_{*\xi,l}$ give contribution, but
now both second term in the r.h.s. of (\ref{la_j}) and $1/n$-order term in the r.h.s. of (\ref{K_exp}) make an impact.

As it was mentioned above, the Laplace-Bertrami operator $\Delta$ on $L_2[S,d\mu]$ has eigenvectors (\ref{phi_j}) with corresponding eigenvalues
$$\lambda_j^*=j(j+3).$$ 
Thus $1-n^{-1}C^*\Delta$ with $C^*=C_*/t_*$ has the same basis of eigenvectors with eigenvalues $1-j(j+3)/t_*W$.

Recall that we are interested in $j\le l-1\sim \log W$ (since $P_l$ is the projection on  $\{\phi_j\}_{j\le l-1}$). Hence, according to (\ref{la_j}) -- (\ref{K_exp}), in the regime  $W^{-1}=C_*n^{-1}$ we can write 
\begin{equation*}
\hat K_{*\xi,l}= P_l\big(1-n^{-1}(C^*\Delta+i\xi\pi \nu\big))P_l+o(n^{-1}),
\end{equation*}
which implies
\begin{equation}\label{oper_lim}
 (\hat K_{*\xi,l}^{n-1}1,1)\to (e^{-C^*\Delta-i\xi\pi\hat\nu}1,1),
\end{equation}
and finishes the proof of Theorem \ref{thm:1}. The detailed proof of (\ref{oper_lim}) is given in Section 6 (see Lemma \ref{l:lim}).

\section{Saddle-point analysis}\label{s:K}
Recall that the stationary  points of the function $\mathcal{F}$ of  (\ref{F_cal})  are defined in (\ref{st_points_1}).

We start the proof from the restriction of the integration with respect to $\bar a_i,\bar a_i'$ by
the neighbourhood of  $a_\pm$.
Set
\begin{align}\notag
\Omega_{+}=&\{x:|x-a_+|\le\log W/W^{1/2}\}, \quad \Omega_{-}=\{x:|x-a_-|\le\log W/W^{1/2}\}, \\ \notag
\widetilde \Omega_{\pm}=&\{a_1,a_1'\in\Omega_+, a_2,a_2'\in\Omega_-\}, \\ \label{Omega}
\widetilde \Omega_{+}=&\{a_1, a_1',a_2,a_2'\in\Omega_+ \},\\ \notag
\widetilde \Omega_{-}=&\{a_1, a_1',a_2,a_2'\in\Omega_- \}
\end{align}
and let $\mathbf{1}_{\widetilde \Omega_{\pm}}$, $\mathbf{1}_{\widetilde\Omega_{+}}$ $\mathbf{1}_{\widetilde\Omega_{-}}$ be indicator functions of the above domains.

\begin{lemma}\label{l:b_K_ab}
Given $A(a,a')$ of (\ref{A_a}), we have
\begin{align}\label{a_K.1}
&\int_{\mathbb{T}\setminus(\Omega_+\cup\Omega_-)}|A(a,a')||da'|\le Ce^{-c\log^2W}.
\end{align}

\end{lemma}
\textit{Proof.} Recall that
\[
a_\pm=e^{\pm i\alpha_0},
\]
and write for the parametrization $a=e^{i\varphi}$, $a'=e^{i\varphi'}$
\begin{align*}
-\Re \Lambda(e^{i\varphi},e^{i\varphi'})=&-\beta(\cos\varphi-\cos\varphi')^2/2+\beta(\sin\varphi-\sin\varphi')^2/2-\frac{\sin^2\varphi+\sin^2\varphi'}{2}\\
 &+\frac{E(\sin\varphi+\sin\varphi')}{2}+\sin^2\alpha_0-E\sin\alpha_0\\
 &=-\beta(\cos\varphi-\cos\varphi')^2/2+\beta(\sin\varphi-\sin\varphi')^2/2\\
 &-(\sin\varphi-\sin\alpha_0)^2/2-(\sin\varphi'-\sin\alpha_0)^2/2\\
&\le\beta(\sin\varphi-\sin\varphi')^2/2-(\sin\varphi-\sin\alpha_0)^2/2-
(\sin\varphi'-\sin\alpha_0)^2/2\\
&\le -(1-2\beta)(\sin\varphi-\sin\alpha_0)^2/2 -(1-2\beta)(\sin\varphi'-\sin\alpha_0)^2/2.
\end{align*}
Here we have used $\sin\alpha_0=E/2$. We have also  for $a'\in \mathbb{T}\setminus(\Omega_+\cup\Omega_-)$
\[
|\sin\varphi'-\sin\alpha_0|\ge C\log W/\sqrt{W}.
\]
Since $\beta<1/4$, this implies (\ref{a_K.1})

$\Box$

Lemma \ref{l:b_K_ab} yields that 
\begin{align}\label{b_out}
\int dQ'd \bar a'(1-\mathbf{1}_{\widetilde\Omega_{\pm}}-\mathbf{1}_{\widetilde\Omega_{+}}-\mathbf{1}_{\widetilde\Omega_{-}})
\|K\|\le e^{-c\log^2W}
\end{align}
Let us prove the following simple proposition
\begin{proposition}\label{p:res} Let the matrix $H(z)$ has the block  form
\begin{equation*}
H(z)=\left(\begin{array}{cc}H_{11}(z)&H_{12}(z)\\H_{21}(z)&H_{22}(z)\end{array}\right).
\end{equation*}
Then 
\begin{align}\label{inv}
G(z)&:=H^{-1}(z)
 =\left(\begin{array}{cc}G_{11}&-G_{11}H_{12}H_{22}^{-1}\\-H_{22}^{-1}H_{21}G_{11}&H_{22}^{-1}+H_{22}^{-1}H_{21}G_{11}H_{12}H_{22}^{-1}\end{array}\right)\\
G_{11}&=(H_{11}-H_{12}H_{22}^{-1}H_{21})^{-1},\
\notag\end{align}
If  $H_{22}^{-1}$ is an analytic function for $|z|>1-\delta$, and $\|H_{22}^{-1}\|\le C$, then
\begin{align}\label{lres}
&\oint_{\omega_A}z^{n-1} (G(z)f, g)dz=\oint_{\omega_A}z^{n-1}(G_{11} f^{(1)}(z), g^{(1)}(z)) dz+ O(e^{-nc})\\
& f^{(1)}(z)=f_0-H_{12}H_{22}^{-1}f_1,\quad  g^{(1)}(z)=g_0-H_{21}^T(H_{22}^T)^{-1}g_1
\notag\end{align}
where $\omega_A=\{z:|z|=1+A/n\}$, $f=(f_0,f_1)$, $g=(g_0,g_1)$ where $f_0$ and $g_0$ are the projection of $f$ and $g$ on the subspace corresponding to
$H_{11}$, while  $f_1$ and $g_1$ are the projection of $f$ and $g$ on the subspace corresponding to
$H_{22}$.
\end{proposition}
\textit{Proof.} Formula (\ref{inv})  is the well-known block matrix inversion formula. Now apply the formula (\ref{inv})  and write
\begin{align*}
&\oint_{\omega_A}z^{n-1} (G(z)f, g)dz=\oint_{\omega_A}z^{n-1}(G_{11} f^{(1)}(z), g^{(1)}(z)) dz+ \oint_{\omega_A} z^{n-1}(H_{22}^{-1}f_1,g_1) dz.
\end{align*}
For the second integral change the integration contour from ${\omega_A}$ to $|z|=1-\delta$. Then the inequality
\[|z|^{n-1}\le (1-\delta)^{n-1}\le Ce^{-nc}\]
yields (\ref{lres}). 

$\square$

Notice that since $\|K\|\le 1$ and $|\mathcal{F}_{n,\xi}|\le 1+C/n$, we can find such $A$ that all eigenvalues of $K_\xi$ lies
inside $\omega_A=\{z:|z|=1+A/n\}$.

Set 
\[H_{11}(z)=H_{11}-z=(\mathbf{1}_{\widetilde\Omega_{\pm}}K_\xi\,\mathbf{1}_{\widetilde\Omega_{\pm}})\oplus(\mathbf{1}_{\widetilde\Omega_{+}}K_\xi\,\mathbf{1}_{\widetilde\Omega_{+}})
\oplus(\mathbf{1}_{\widetilde\Omega_{-}}K_\xi\,\mathbf{1}_{\widetilde\Omega_{-}})-z=K_{\xi,\pm}\oplus K_{\xi, +}\oplus K_{\xi, -}-z.\]
Then (\ref{b_out}) yields
\[\|H_{22}\|+\|H_{12}\|+\|H_{21}\|\le Ce^{-c\log^2W}.\]
Therefore for any $|z|>\frac{1}{2}$
\[\|H_{12}(H_{22}-z)^{-1}H_{21}\|\le  Ce^{-c\log^2W}.\]
Moreover, it will be proven below that 
\begin{equation*}
\|(H_{11}-z)^{-1}\|\le Cn, \quad z\in\omega_A,
\end{equation*}
and so for $G_{11}$ of (\ref{inv}) we have
\[
\|G_{11}-(H_{11}-z)^{-1}\|\le e^{-c\log^2W/2}.
\]
Here we have used $W\ge n^\varepsilon$. Thus
we obtain by Proposition \ref{p:res}
\begin{align}
 \oint_{\omega_A} z^{n-1}(G_\xi(z)f,g)dz=\oint_{\omega_A}  z^{n-1}((H_{11}-z)^{-1}f,g)dz+O(e^{-c\log^2W/2})+O(e^{-nc_1}),
\label{repr1} \end{align}
where $G_\xi(z)$ is a resolvent of $K_\xi$ (see (\ref{res_rep})). 
In view of  the block structure of $H_{11}$, its resolvent  also has a block structure, hence 
\begin{align}\notag
 \oint_{\omega_A} z^{n-1}(G_\xi(z)f,g)dz=&\oint_{\omega_A} z^{n-1}
(G_{\xi,\pm}(z)f_{\pm},g_\pm)dz+\oint_{\omega_A}  z^{n-1}(G_{\xi,+}(z)f_{+},g_+)dz
\\&+\oint_{\omega_A}  z^{n-1}(G_{\xi,-}(z)f_{-},g_-)dz
=I_{\xi,\pm}+I_{\xi,+}+I_{\xi,-},
\label{repr2}\end{align}
where 
\begin{align*}
G_{\xi, \pm}=&(K_{\xi,\pm}-z)^{-1},\quad 
 G_{\xi,+}(z)=(K_{\xi,+}-z)^{-1},\quad
G_{\xi,-}(z)=(K_{\xi, -}-z)^{-1} 
\end{align*}
and $f_{\pm},f_+,f_-$, $g_{\pm},g_+,g_-$ are projections of $f$ and $g$ onto the subspaces corresponding to
$K_{\xi,\pm},\,K_{\xi,+},\,K_{\xi,-}.$ One can perform similar analysis for $K$ instead of $K_\xi$ and define $I_\pm$, $I_+$,
and $I_-$.

In the next sections we are going to study each integral $I_{\xi,\pm}$, $I_{\xi,+}$, and $I_{\xi,-}$ separately.
It will be shown below (see Section 7) that that $I_{\xi,+}$ and $I_{\xi,-}$ are exponentially small comparable to $I_{\xi,\pm}$,
so the main task is to study $I_{\xi,\pm}$.

\section{Analysis of $I_{\xi,\pm}$}
As was mentioned in Section 4, to analyze $K_\pm$ and $K_{\xi,\pm}$ we are going to use the polar decomposition 
(\ref{rep_2}) -- (\ref{phi_0}).

We start with the analysis of operator $A_a$ of (\ref{A_a}) in the domain $\widetilde\Omega_\pm$  of (\ref{Omega}).

To this end, we are going to consider quadratic approximation of $A(a,a')$ defined in (\ref{A_a}).
Make a change of variables
 \begin{align}\label{ch_ab}
 &a_{1i}=a_+(1+i\theta_+\widetilde{a}_{1i}/\sqrt{W}),\quad a_{2i}=a_-(1+i\theta_-\widetilde{a}_{2i}/\sqrt{W}),
 \end{align}
where $\theta_\pm$ are some complex constants with $|\theta_\pm|=1$ which will be determined later (see (\ref{kappa})). 
Notice that the Jacobian of (\ref{ch_ab}) is a constant depending on $n,W$ but not on $\xi$, thus it does not give contribution
to $\tilde C'_{n,W}$ (see (\ref{F_rep1})).
Define
\begin{align}\label{A_pm}
A^+(\tilde a,\tilde a')&=1_{\Omega_+}A\Big(a_+(1+i\theta_+\tilde a/\sqrt{W}), a_+(1+i\theta_+\tilde a'/\sqrt{W})\Big) 1_{\Omega_+},\\
A^-(\tilde a,\tilde a')&=1_{\Omega_-}A\Big(a_-(1+i\theta_-\tilde a/\sqrt{W}), a_-(1+i\theta_-\tilde a'/\sqrt{W})\Big) 1_{\Omega_-}. \notag
\end{align}
Then
\begin{multline}\label{K_pm}
K_{\xi,\pm}(a_1,a_2,Q;a'_1,a'_2,Q')\\
=A^+(\tilde a_1,\tilde a'_1)A^-(\tilde a_2,\tilde a'_2)K_*(t,Q,Q')e^{-\frac{i\xi\pi}{n} \big(\nu(a_1-a_2,Q)+\nu(a_1'-a_2',Q')\big)}.
\end{multline}

Since $\varphi''_0(a_+)=c_+$ (see (\ref{phi_0}) and (\ref{c_pm})), it is easy to see that 
the kernel $A_a^+$  of (\ref{A_pm}) takes the form
\begin{align}\label{K_a*}
 A^+(\tilde a,\tilde a')&=A^+_*(\tilde a,\tilde a')(1+W^{-1/2}\hat p_+(\tilde a))(1+W^{-1/2}\hat p_+(\tilde a'))+O(e^{-c\log^2W}),\, \\
A^+_*(\tilde a,\tilde a')&=\dfrac{a_+\theta_+}{\sqrt{2\pi}}\exp\Big\{(a_+\theta_+)^2\big[\beta (\tilde a-\tilde a')^2/2-c_+\tilde a^2/4-c_+(\tilde a')^2/4\big]\Big\}\notag \\
\hat p_+(\tilde a)&=ic_{3+}\tilde a^3-c_{4+}\tilde a^4W^{-1/2}-ic_{5+}\tilde a^5W^{-1}+\dots
\notag\end{align}
where the coefficients  $c_{3+},c_{4+},\dots $ are expressed in terms of the derivatives of $\varphi_0$ at $a_+$.
Similarly $A^-$ of (\ref{A_pm}) can be approximated via $A^-_*$ defined similarly to $A^+_*$ in
~(\ref{K_a*}).

It is easy to check that for $\beta<1/4$ the real parts of the eigenvalues $\alpha_{1,+}$, $\alpha_{2,+}$ of the quadratic form
\begin{align*}
\left(\begin{array}{cc}
a_+^2(\frac{c_+}{2}-\beta)&a_+^2\beta\\
a_+^2\beta& a_+^2(\frac{c_+}{2}-\beta)
\end{array}\right)
\end{align*}
in the exponent of $A^+_*$ of (\ref{K_a*}) are positive. Same is true for $A^-_*$. 
Denote 
\begin{align}
&\theta_{\pm} =(|\kappa_{\pm}|/\kappa_{\pm})^{1/2},\quad \kappa_{\pm}=(\alpha_{1,\pm}\alpha_{2,\pm})^{1/2}=a_\pm^2\big( c_{\pm}^2/4-\beta c_{\pm}\big)^{1/2},\label{kappa}
 \end{align}
with $c_\pm$ of (\ref{c_pm}). Notice that $\theta_\pm$ is defined in such a way that
\begin{align*}
\Re (\theta_\pm^2\alpha_{1,\pm})>0, \quad \Re (\theta_\pm^2\alpha_{2,\pm})>0.
\end{align*}
Now introduce the orthonormal bases
\begin{align}\label{psi_k+}
\psi_k^\pm(\tilde a)=|\kappa_\pm|^{1/4}H_k(|\kappa_\pm|^{1/2}\tilde a) e^{-|\kappa_\pm|\tilde a^2/2},
\end{align}
where $\{H_k(x)\}$ are   Hermit polynomials which are orthonormal with the weight $e^{-x^2}$:
\[
H_k(x)=(2^{k-1/2}k!\sqrt{2\pi} )^{-1/2}e^{x^2}(\frac{d}{dx})^ke^{-x^2}.
\]
Below we will need the following lemma 
\begin{lemma}\label{l:A_ab}

\begin{enumerate}
\item[(i)] Let $\kappa_{+},\kappa_-$ be defined as in (\ref{kappa}). 
Then the matrices of the operators $A_*^+$ and $A_*^-$ are diagonal in the basis $\{\psi_k^+\}$ and $\{\psi_k^-\}$
and the corresponding eigenvalues have the form
\begin{align}\label{la_k}
\lambda_k^\pm&=\lambda_k(A_*^\pm)=\lambda_0^\pm \cdot q_\pm^k,\quad k=0,1,2\dots
\end{align}
with
\begin{align}\label{lam_0}
&\lambda_0^\pm=(\kappa_\pm/a_\pm^2+c_\pm/2-\beta)^{-1/2},\\ \notag
&q_\pm=\frac{\beta}{\kappa_\pm/a_\pm^2+c_\pm/2-\beta}, 
\quad |q_\pm|<1.
\end{align}
Notice that $|q_\pm|<1$ implies
\begin{equation}\label{lam_bound}
|\lambda_0^\pm|\le \beta^{-1/2}.
\end{equation}
The matrices of operators $A^+$ and $A^-$ of (\ref{A_pm}) have the form
\begin{align}\label{A_matr}
(A^\pm)_{k,k}&=\lambda_0^\pm\cdot q_\pm^k+O(1/W),\\ \notag
(A^\pm)_{k,k'}&=O(W^{-1/2})(\delta_{|k-k'|,1}+\delta_{|k-k'|,3})\\ \notag
&+O(W^{-1})\delta_{|k-k'|,2}+O(W^{-(|k-k'|-3)/2}), \quad k\ne k'.
\end{align}
\item[(ii)] The eigenvalues of operator 
\begin{equation}\label{Apm}
A_{\pm}=1_{\tilde\Omega_\pm}\big(\lambda_0(t)A_a\big) 1_{\tilde\Omega_\pm}
\end{equation} 
 are $\lambda_0^+\lambda_0^{-}q_+^kq_-^l+O(1/W)$, $k,l=0,1,..$
and they are solutions of  the equation
\begin{align}\label{eq_A}
(A_{\pm})_{0,0}-z-(A_\pm)^{(12)}((A_\pm)^{(22)}-z)^{-1}(A_\pm)^{(21)}=0,
\end{align}
where 
\[
A_\pm=\left(\begin{array}{cc}
A_{00}&A^{(12)}\\
A^{(21)}&A^{(22)}
\end{array}\right)
\]
according to the decomposition $\{\psi_{k_1}^+\psi_{k_2}^-\}=\{\psi_{0}^+\psi_{0}^-\}\oplus \{\psi_{k_1}^+\psi_{k_2}^-\}_{\bar k\ne 0}$ with $\bar k=(k_1,k_2)$.
Here $\lambda_0(t)$ is the top eigenvalue of $K_*(t,Q,Q')$ (see (\ref{l_j})).

The top eigenvalue of $K_{\pm}$ has the form 
\begin{align*}
\lambda_0(K_\pm)=\lambda_0(A_\pm)=\lambda_0^+\lambda_0^{-}+O(1/W).
\end{align*}
\end{enumerate}
\end{lemma}

\textit{Proof.} To simplify formulas, we consider the kernel (see (\ref{K_a*}) -- (\ref{kappa}))
\begin{align*}
M(x,y)=a_+(2\pi)^{-1/2}e^{-(\mathcal{A}x,x)/2},\,\bar x=(x,y),\quad \mathcal{A}=\left(\begin{array}{cc}\mu&\nu\\ \nu&\mu \end{array}\right),\,\lambda_{\pm}=\mu\pm \nu,\,\Re\lambda_\pm>0.
\end{align*}
Then, taking $\kappa=\sqrt{\mu^2-\nu^2}=\sqrt{\lambda_+\lambda_-}$, we obtain that 
\begin{align*}
a_+(2\pi)^{-1/2}\int e^{-(\mathcal{A}x,x)/2+\kappa y^2/2}&(\frac{d}{dy})^ke^{-\kappa y^2}\ dy\\
&=q^k \cdot a_+(\mu+\kappa)^{-1/2}e^{\kappa x^2/2}(\frac{d}{dy})^ke^{-\kappa x^2},
\quad q=\frac{\nu}{\mu+\kappa},
\end{align*}
so $e^{\kappa y^2/2}(\frac{d}{dy})^ke^{-\kappa y^2}$, $k=0,1,\ldots $ are the eigenvectors of $M$.
Since  $M$ is compact, we  have  $|q|< 1$. Notice also that
\[
a_+(\mu+\kappa_\pm)^{-1/2}=\lambda_0^\pm.
\]
Now if we change the variables
 \[x_1=\theta x,\, y_1=\theta y,\quad \theta =e^{-i(\mathrm{arg}\,\lambda_++\mathrm{arg}\,\lambda_-)/4}=e^{-i\mathrm{arg}\,\kappa/2}, \]
 then
for the new matrix $\widetilde A=\theta^2A$ has eigenvalues $\theta^2\lambda_+, \theta^2\lambda_-$, whose real parts  are still positive,
 $\widetilde\kappa=|\kappa|$, and $\widetilde q=q$. This finishes the proof of  (\ref{la_k}) -- (\ref{lam_0}).
 
 Formula (\ref{A_matr}) follows directly from (\ref{K_a*}) and the fact that the Gaussian integral of $x^{2k+1}$ is zero,
 and it immediately gives the statement about eigenvalues of $A_\pm$ (it is easy to see that $\lambda_0(t)$ does not change anything
 since it has only $\tilde a/W^{3/2}$ and $\tilde a'/W^{3/2}$).

Equation (\ref{eq_A}) can be obtained from the standard Schur inversion formula. The rest of part (ii) follows directly from (i)
and Proposition \ref{p:K(U)}.


 $\Box$ 
 
Now we are going to normalize  $K_\pm$, $K_{\xi,\pm}$ by $\lambda_0(K_\pm)$:
\begin{equation}\label{K_hat}
\hat K_\pm=\lambda_0(K_\pm)^{-1}K_{\pm},\quad \hat K_{\xi,\pm}=\lambda_0(K_\pm)^{-1}K_{\xi,\pm}
\end{equation}
with $K_{\pm,\xi}$ of (\ref{K_pm}). Notice that
\begin{equation}\label{K*_hat}
\hat K_\pm=\hat A_\pm\cdot \hat K_*
\end{equation}
where
\begin{equation*}
\hat A_\pm= (\lambda_0(A_\pm))^{-1} A_\pm, \quad \hat K_*(t,Q,Q')=(\lambda_0(t))^{-1}K_*(t,Q,Q'),
\end{equation*}
so both top eigenvalues of $\hat A_\pm$, $\hat K_*$ is 1, and
\begin{equation}\label{lam_hat}
\hat \lambda_j(\hat K_*)=1-\dfrac{j(j+3)}{tW}+O((j^2/tW)^2), \quad j=1,2,\ldots.
\end{equation}
Therefore, it is easy to see that all eigenvalues of $\hat K_\pm$, $\hat K_{\xi,\pm}$ lies inside $\omega_A=\{z: |z|=1+A/n\}$. Thus
we get 
 \begin{equation*}
 I_{\pm,\xi}=-2\pi i(K_{\xi,\pm}^{n-1}f,g)=-2\pi i \cdot \lambda_0(K_\pm)^{n-1}(\hat K_{\xi,\pm}^{n-1}f,g)=
 \lambda_0(K_\pm)^{n-1}\int_{\omega_A} z^{n-1} (\hat G_{\xi}(z)f,g) d z,
 \end{equation*}
 where
\begin{equation*}
\hat G_{\xi}(z)=(\hat K_{\xi,\pm}-z)^{-1}.
\end{equation*} 
Similarly we can rewrite $I_{\pm}$. 

Consider the matrix of  $\hat K_{\xi,\pm}$  in the basis
\begin{equation}\label{basis}
\Psi_{\bar k, j}(\tilde a_1,\tilde a_2, Q)=\psi^+_{k_1}(\tilde a_1)\psi^-_{k_2}(\tilde a_2)\phi_j(Q), \quad k_1,k_2,j\ge 0,
\end{equation}
 with $\psi_{k}^\pm$ of (\ref{psi_k+}), and $\phi_j$ of (\ref{phi_j}). Let $\mathcal{H}_1=\{\Psi_{\bar k,j}\}_{ j\le l-1}$ and
 \begin{equation}\label{decomp}
 L_2(\mathbb{R}^2)\times L_2(\mathring{Sp}(2), d\mu(Q))=\mathcal{H}_1\oplus \mathcal{H}_2,
 \end{equation}
 and write 
 \begin{equation}\label{K_block}
  \hat K_{\pm}=\left(
 \begin{array}{cc}
 \hat K^{(11)}&\hat K^{(12)}\\
 \hat K^{(21)}&\hat K^{(22)}
 \end{array}\right), \quad \hat K_{\xi,\pm}=\left(
 \begin{array}{cc}
 \hat K^{(11)}_\xi&\hat K^{(12)}_\xi\\
 \hat K^{(21)}_\xi&\hat K^{(22)}_\xi
 \end{array}\right)
 \end{equation}
 according to this decomposition. We will need the following simple lemma
 \begin{lemma}\label{l:decomp1}
 Given decomposition (\ref{K_block}), we have
 \begin{align}\label{K_12}
 &\hat K^{(12)}=\hat K^{(21)}=0, \quad \|\hat K^{(12)}_\xi\|\le \dfrac{C}{n},\quad  \|\hat K^{(21)}_\xi\|\le \dfrac{C}{n};
 \end{align}
 and for $|z|\ge 1+A/n$ with big enough $A$ we have
 \begin{align}\label{K_11}
 &\|(\hat K^{(11)}_\xi-z)^{-1}\|\le Cn,\\ \label{K_22}
 &\|(\hat K^{(22)}_\xi-z)^{-1}\|\le CW/l^2,
 \end{align}
 and same is valid for $\hat K$.
 \end{lemma}
 \noindent{\it Proof of Lemma \ref{l:decomp1}.}
 The bound (\ref{K_12}) follows follows from the  block-diagonal structure of $\hat K_\pm$  with respect to the basis (\ref{basis}) (see (\ref{block})), 
and the fact that $\hat K_{\pm,\xi}$ is $1/n$-order perturbation of $\hat K_\pm$.

The bound (\ref{K_11}) follows from
\[
\|\hat K^{(11)}_\xi\|\le 1+c/n, 
\]
since for big enough $A$
\[
\|(\hat K^{(11)}_\xi-z)^{-1}\|\le |z|^{-1}\sum\limits_{k=0}^\infty\left(\dfrac{\|K^{(11)}_\xi\|}{|z|}\right)^k\le Cn.
\]
Similarly, according to (\ref{l_j}) -- (\ref{block}), we get
\[
\|\hat K^{(22)}\|\le 1-\dfrac{Cl(l+3)}{W}
\]
and, since $l^2/W\sim \log^2n/n$ for $W\ge Cn$ and $l^2/W\gg n^{-1}$ for $W\ll n$,
\[
\|\hat K^{(22)}_\xi\|\le 1-\dfrac{Cl^2}{W},
\]
which implies (\ref{K_22}). 
 $\Box$
 
The next step is to prove that we can consider only the upper-left block $K^{(11)}_\xi$ of $K_\xi$ (see (\ref{K_block})). More precisely, we are going to prove
\begin{lemma}\label{l:K_tr}
We have
\begin{equation*}
\int\limits_{\omega_A} z^{n-1}(\hat G_\xi (z) f, \bar f) dz=\int\limits_{\omega_A} z^{n-1}( \hat G_{1,\xi}(z)\,f_{1}, \bar f_{1}) dz + O\Big (\dfrac{ W\log n}{l^2 n}\Big),
\end{equation*} 
where
\begin{equation*}
\hat G_{1,\xi}(z)=(\hat K^{(11)}_\xi-z)^{-1},
\end{equation*}
and we decomposed  $f=(f_1,f_2)$  with respect to decomposition (\ref{decomp}).
Notice that 
\[
\dfrac{ W\log n}{l^2 n}\le \dfrac{1}{\log n }.
\]
\end{lemma}
\textit{Proof.}
Using the well-known Schur inversion formula we get
\begin{align*}
(\hat K_\xi-z)^{-1}=\left(\begin{array}{cc} \hat G_\xi^{(11)}&-\hat G_\xi^{(11)}\hat K^{(12)}_\xi \hat G_{2,\xi}\\ -\hat G_{2,\xi} \hat K^{(21)}_\xi \hat G_\xi^{(11)} & \hat G_{2,\xi}+
\hat G_{2,\xi} \hat K^{(21)}_\xi \hat G_\xi^{(11)} \hat K^{(12)}_\xi \hat G_{2,\xi}\end{array}\right),
\end{align*}
where
\begin{align*}
&\hat G_{2,\xi}(z)=(\hat K^{(22)}_\xi-z)^{-1},\\
&\hat G_\xi^{(11)} =(\hat K^{(11)}_\xi-z-\hat K^{(12)}_\xi \hat G_{2,\xi} \hat K^{(21)}_\xi)^{-1}=(1-\hat G_{1,\xi}\hat K^{(12)}_\xi \hat G_{2,\xi} \hat K^{(21)}_\xi)^{-1} \hat G_{1,\xi}.
\end{align*}
Thus
\begin{align}\label{sum_int}
&\int_{\omega_A} z^{n-1} ((K_\xi-z)^{-1}f,\bar f) dz=\int_{\omega_A} z^{n-1} (G^{(11)}_\xi f_{1},\bar f_{1}) dz-\int_{\omega_A} z^{n-1} (G_\xi^{(11)}K^{(12)}_\xi G_{2,\xi} f_2,\bar f_1) dz\\ \notag
&-\int_{\omega_A} z^{n-1} (G_{2,\xi} K^{(21)}_\xi G_\xi^{(11)} f_1,\bar f_2) dz+\int_{\omega_A} z^{n-1} ((G_{2,\xi}+
G_{2,\xi} K^{(21)}_\xi G_\xi^{(11)} K^{(12)}_\xi G_{2,\xi}) f_2,\bar f_2) dz.
\end{align}
Denoting
\begin{equation*}
R=(1- \hat G_{1,\xi}\hat K^{(12)}_\xi \hat G_{2,\xi} \hat K^{(21)}_\xi)^{-1},
\end{equation*}
we get
\begin{equation*}
\hat G_\xi^{(11)} =R \hat G_{1,\xi}.
\end{equation*}
According to (\ref{K_12}) -- (\ref{K_22}), we obtain
\begin{align*}
\|\hat G_{1,\xi}\hat K^{(12)}_\xi \hat G_{2,\xi} \hat K^{(21)}_\xi\|\le Cn\cdot \dfrac{1}{n^2}\cdot \dfrac{W}{l^2}=\dfrac{CW}{l^2 n}.
\end{align*}
Therefore
\begin{align*}
\|1-R\|\le \dfrac{CW}{l^2n},
\end{align*}
which together with (\ref{K_11}) imply
\begin{align*}
\Big|\int_{\omega_A}z^{n-1}\Big(\big(\hat G_\xi^{(11)}-\hat G_{1,\xi}\big)f_1,\bar f_1\Big)\Big|&=\Big|\int_{\omega_A}z^{n-1}\Big((1-R)G_{1,\xi}\big)f_1,\bar f_1\Big)\Big|\\
&\le C\|1-R\|\cdot \|f_1\|^2\cdot \int_{\omega_A}\dfrac{|dz|}{|z-1|}\le \dfrac{CW\log n}{l^2n}.
\end{align*}
It is easy to see also that 
\[
\|f_2\|\le C/n,
\]
and because of the consideration above
\begin{align*}
&\|\hat G_{2,\xi}+\hat G_{2,\xi} \hat K^{(21)}_\xi \hat G_\xi^{(11)} \hat K^{(12)}_\xi \hat G_{2,\xi}\|\le CW/l^2,\\
&\|\hat G_{2,\xi} \hat K^{(21)}_\xi \hat G_\xi^{(11)} \|\le CW/l^2,\quad \| \hat G_\xi^{(11)} \hat K^{(12)}_\xi  \hat G_{2,\xi}\|\le CW/l^2,
\end{align*}
so other terms in (\ref{sum_int}) are also small.

$\Box$

The next step is to show that we can consider only the projection of  $\hat K_{\xi}^{(11)}$, $\hat K^{(11)}$ on a linear span of of $\{\Psi_{0,j}\}_{j\le l}$ (see (\ref{basis})).
We prove
\begin{lemma}
Let $P_l$ be the projection on $\{\phi_j\}_{j=0}^{l-1}$ of (\ref{phi_j}), $\Delta_l=P_l\Delta P_l$, and $\hat\nu_l=P_l\hat\nu P_l$ with $\hat \nu$ defined in (\ref{nu}).
Then
\begin{align*}
&\int\limits_{\omega_A} z^{n-1}( \hat G_{1,\xi}(z)\,f_{1}, \bar f_{1}) dz\\
&=\int\limits_{\omega_A} \zeta^{n-1}\Big (\Big(P_l-\dfrac{1}{t_*W}\Delta_l-\dfrac{i\pi\xi}{n}\hat\nu_l-\zeta+
O\Big(\dfrac{(l-1)^4}{W^2}\Big)\Big)^{-1}f_{0}, \bar f_{0}\Big) dz+O\Big(\dfrac{(l-1)^2n}{W^{3/2}}\Big)+O\Big(\dfrac{1}{W^{1/2}}\Big),
\end{align*}
where $O(x)$ is an operator whose norm is bounded by $Cx$ which does not depend on $\zeta$,
and
\[
f_0=(f,\Psi_{\bar 0}).
\]
Recall $l=1$ for $W\ll n$ and $(l-1)^2n/W^{3/2}\sim \log^2n/W^{1/2}$ for $W\ge Cn$, and $t_*=(2\pi\rho(E))^2$.
Similar formula is true for $\hat G_{1}$ (i.e. for $\hat K^{(11)}$ instead of $\hat K_{\xi}^{(11)}$).
\end{lemma}
\textit{Proof.}
Now write $\hat K_{\xi}^{(11)}-z$, $\hat K^{(11)}-z$ in the block form
\begin{equation*}
\hat K^{(11)}-z=\left(\begin{array}{cc}
M_{1}&M_{12}\\
M_{21}&M_{2}
\end{array}\right),\quad \hat K_{\xi}^{(11)}-z=\left(\begin{array}{cc}
M_{1,\xi}&M_{12,\xi}\\
M_{21,\xi}&M_{2,\xi}
\end{array}\right)
\end{equation*}
according to decomposition
\[
\mathcal{H}_1=\mathcal{M}_1\oplus \mathcal{M}_2,
\]
where $\mathcal{M}_1$ is a linear span of $\Psi^{(l)}_0=\{\Psi_{0,j}\}_{j\le l}$ (see (\ref{basis})).

Set
\begin{equation*}
G^0_{1,\xi}(z)=(M_{11,\xi}-M_{12,\xi}M_{22,\xi}^{-1}M_{21,\xi})^{-1}\quad G^0_{1}(z)=(M_{11}-M_{12}M_{22}^{-1}M_{21})^{-1}
\end{equation*}
Then, using Proposition \ref{p:res}, we get
\begin{align*}
&\oint_{\omega_A}z^{n-1} (G(z)f_1, \bar f_1)dz=\oint_{\omega_A}z^{n-1}(G^0_{1} f^{(1)}(z), g^{(1)}(z)) dz+ O(e^{-nc}),\\
&\oint_{\omega_A}z^{n-1} (G_\xi(z)f_1, \bar f_1)dz=\oint_{\omega_A}z^{n-1}(G^0_{1,\xi} f^{(1)}(z), g^{(1)}(z)) dz+ O(e^{-nc}),
\notag\end{align*}
where $f^{(1)}$, $g^{(1)}$ are defined as in (\ref{lres}). 
Recall that 
\begin{align*}
\hat K^{(11)}-z=(\hat AP_l\hat K_*P_l\Psi_{0}^{(l)},\Psi_0^{(l)})-z.
\end{align*}
Set
\begin{align*}
P_l\hat K_*P_l=P_l-\tilde{K}_l.
\end{align*}
Then $\tilde{K}_l$, according to (\ref{lam_hat}), is a diagonal matrix with eigenvalues 
\begin{equation*}
\tilde \lambda_{j}(t)=j(j+3)/tW+O((j^2/tW)^2), \quad j=0,\ldots,l-1.
\end{equation*} 
Since (\ref{t}) and (\ref{ch_ab}) implies
\[
t=\Big(a_+-a_-+\dfrac{i\theta_+a_+\tilde a_1-i\theta_-a_-\tilde a_2)}{\sqrt{W}}\Big)\Big(a_+-a_-+\dfrac{i\theta_+a_+\tilde a'_1-i\theta_-a_-\tilde a'_2)}{\sqrt{W}}\Big),
\]
$\tilde{K}_l$ can be rewritten as
\begin{align}\label{K_*-exp}
\tilde{K}_l=\Delta_l/t_*W +O_a\Big(\dfrac{(l-1)^2}{W^{3/2}}\Big)+O\Big(\dfrac{(l-1)^4}{W^{2}}\Big)
\end{align}
with $t_*=(a_+-a_-)^2=(2\pi\rho(E))^2$. 
Here  $O_{a}(X)$ is an diagonal in $\{\phi_j\}_{j=0}^{l-1}$ operator of the type $O(X)$ whose eigenvalues are
linear in $a, a'$.

Now, since, according to Lemma \ref{l:A_ab}, $\hat A_{00}=1+O(1/W)$, substituting (\ref{K_*-exp}), we get
\[
(\hat A\tilde K_l\Psi^{(l)}_{0},\Psi^{(l)}_0)_{jj}=( \tilde \lambda_{j}(t)\hat A \psi_0^+\psi_0^-, \psi_0^+\psi_0^-)=\dfrac{j(j+3)}{t_*W}\cdot \Big(1+O\Big(\dfrac{j^2}{W}\Big)\Big)
\]
Therefore,  
\begin{align*}
\hat K^{(11)}-z=\hat A_{00}P_l-z- \Delta_l/t_*W +O((l-1)^4/W^2).
\end{align*}
Similarly
\begin{align*}
&M_{12}= \hat A_{12}\otimes P_l+O\Big(\frac{(l-1)^2}{W^{3/2}}\Big),\\
& M_{21}=\hat A_{21}\otimes P_l+O\Big(\frac{(l-1)^2}{W^{3/2}}\Big),\\
&M_{22}=\hat A_{22}\otimes P_l-z+O\Big(\frac{(l-1)^2}{W}\Big).
\end{align*}
Notice also that because of Lemma \ref{l:A_ab} $$\|\hat A_{12}\|\le W^{-1/2},\quad \|\hat A_{21}\|\le W^{-1/2},\quad \|(\hat A_{22}-z)^{-1}\|\le C.$$ Hence
\begin{align*}\notag
&M_{11}-M_{12}M_{22}^{-1}M_{21}=A_{00}P_l-\Delta_l/t_*W-z-\hat A_{12}(\hat A_{22}-z)^{-1}\hat A_{21}P_l+O_z\Big(\frac{(l-1)^2}{W^{2}}\Big)\\
&+O\Big(\frac{(l-1)^4}{W^{2}}\Big)
=P_l-\Delta_l/t_*W -z -W^{-1} g_1(z) P_l+O\Big(\frac{(l-1)^4}{W^{2}}\Big),
\end{align*}
where
\begin{align*}
g_1(z)=\hat A_{00}-1-\hat A_{12}(\hat A_{22}-z)^{-1}\hat A_{21}+O_z((l-1)^2/W)-O_*((l-1)^2/W)
\end{align*}
 is analytic and bounded in $\{z: |z-1|<\delta\}$ for small enough $\delta$ (recall $l^2/W\le \log^2n/n$ and $\|\hat A^{22}\|\le |q_\pm|<1$ with
 $q_\pm$ of (\ref{lam_0}) according to Lemma \ref{l:A_ab}). Here $O_z(\cdot)$ is operator of type $O(\cdot)$ which may depend on $z$, and $O_*(\cdot)$ is operator
 $O_z(\cdot)$ with substitution $z=1$.  Lemma \ref{l:A_ab} implies also
 \begin{align}\label{g_1(0)} 
g_1(1)=0.
\end{align}
Now set 
\begin{align}\label{zeta}
\zeta(z)=z+W^{-1} g_1(z).
\end{align}
Since $g_1(z)$ has a bounded derivative in $\{z: |z-1|<\delta\}$, we get
\begin{align*}
\zeta'(z)=1+O(1/W),
\end{align*}
the Implicit function theorem implies that there exists the inverse function $z(\zeta)$ with a derivative of order $1+O(1/W)$.
In addition, by (\ref{g_1(0)}), $\zeta(1)=1$, so the image of $\{z: |z-1|<\delta\}$ lies in $\{\zeta:\delta/2<|\zeta-1|<2\delta\}$, and 
it is easy to show that
\[
z(\zeta)=\zeta+W^{-1}\tilde g_1(\zeta),
\]
where $\tilde g_1$ is a bounded analytic in $\{\zeta:|\zeta-1|<2\delta\}$, and $\tilde g_1(1)=0$.

Now we consider the contour $\tilde\omega_A=\{z: \hbox{dist}\{z; [1-C(l-1)(l+2)/W; 1]\}\le A/n\}$ and the contour $\mathcal{L}_2=\{|z|\le \dfrac{|q_\pm|+1}2<1\}$ with $q_\pm$
of (\ref{lam_0}). It is easy to see
that $\tilde\omega_A\cup \mathcal{L}_2$ encircle all the eigenvalues of $\hat K^{(11)}$, $\hat K^{(11)}_\xi$ (see (\ref{l_j}) -- (\ref{block}) and Lemma \ref{l:A_ab}).
But for $z\in\mathcal{L}_2$
\[
|z|^{n-1}\le e^{-cn},
\]
so the contribution of the integral over $\mathcal{L}_2$ is small, and we need to consider integral over $\tilde\omega_A$ only.
It follows from 
(\ref{g_1(0)}) -- (\ref{zeta}) that and consideration above that  $\zeta(z), z\in \tilde\omega_A$ will be inside
$\tilde\omega_{2A}$, and so, since $l=1$ for $W\ll n$ and $l^2/W\sim \log^2n/n$ for $W\ge Cn$, we get 
\[
z(\zeta)=\zeta+O((l-1)^2/W^2)+O(1/nW),
\]
hence
\begin{align}\label{zn}
z^{n-1}=\left\{\begin{array}{ll}
\zeta^{n-1}+O(\log^2n /W),& W\ge Cn,\\
\zeta^{n-1}+O(1 /W),& W\ll n.
\end{array}\right.
\end{align}
Notice also that for $z\in \tilde\omega_A$
\[
\|(P_l-\Delta_l/t_*W-\zeta(z))^{-1}\|\le Cn, 
\]
thus
\begin{align*}
&\|G_1^0\|=\|(P_l-\Delta_l/t_*W-\zeta(z)+O((l-1)^4/W^2))^{-1}\|\le \|(P_l-\Delta_l/t_*W-\zeta(z))^{-1}\|\\
&\times\Big\|\Big(1+O((l-1)^4/W^2)\cdot (P_l-\Delta_l/t_*W-\zeta(z))^{-1}\Big)^{-1}\Big\|\le Cn.
\end{align*}
Hence, recalling $l=1$ and $\|\tilde\omega_A\|=C/n$ for $W\ll n$ and $|\tilde\omega_A|\le C (l-1)^2/W$ for $W\ge Cn$, and  $\| f^{(1)}(z)-f_0\|\le C/\sqrt{W}$, we obtain
\begin{multline*}
\oint_{\tilde \omega_A}z^{n-1}(G_1^0(z) f^{(1)}(z), g^{(1)}(z)) dz\\=\oint_{\tilde\omega_A}z^{n-1}(G_1^0(z) f_0, \bar f_0) dz+O((l-1)^2n/W^{3/2})+O(1/W^{1/2}),
\end{multline*}
that according to (\ref{zn}) can be further transformed as
\begin{align*}
&\oint_{\tilde\omega_A}z^{n-1}(G_1^0(z) f_0, \bar f_0) dz\\
&=\oint_{\zeta(\tilde\omega_A)}\zeta^{n-1}((P_l-\Delta_l/t_*W-\zeta+O((l-1)^4/W^2))^{-1} f_0, \bar f_0) d\zeta+O\Big(\dfrac{(l-1)^4n^2}{W^3}\Big)+
O\Big(\dfrac{1}{W}\Big),
\end{align*}
and the contour now can be changed back to $\omega_A$ (notice $l^4n^2/W^3=\log^4 n/W$, $l^4/W^2=\log^2n/Wn$ for $W\ge Cn$).

In order to perform the same analysis for $\hat K_\xi^{(11)}$ notice that
\begin{align*}
&\|M_{12}\|\le C/\sqrt{W}, \quad M_{12,\xi}=M_{12}+O\Big(\frac{1}{n\sqrt{W}}\Big);\\
&\|M_{21}\|\le C/\sqrt{W}, \quad M_{21,\xi}=M_{21}+O\Big(\frac{1}{n\sqrt{W}}\Big);\\
&\|M_{22}^{-1}\|\le C,\quad \quad \quad M_{22,\xi}=M_{22}+O\Big(\frac{1}{n}\Big),
\end{align*}
and 
\begin{align*}
M_{11,\xi}=(AP_lF_{n,\xi}K_*F_{n,\xi}P_l\Psi_{0},\Psi_0)-z=M_{11}-A_{00}\cdot \dfrac{i\pi\xi}{n} P_l \hat{\nu} P_l+O\Big(\frac{1}{n\sqrt{W}}\Big).
\end{align*}
Thus, since $A_{00}=1+O(1/W)$, we have
\begin{align*}
M_{1,\xi}-M_{12,\xi}M_{22,\xi}^{-1}M_{21,\xi}=M_{11}-M_{12}M_{22}^{-1}M_{21}-\dfrac{i\pi\xi}{n} P_l \hat{\nu} P_l+O\Big(\frac{1}{n\sqrt{W}}\Big),
\end{align*}
and hence we can apply same consideration as above.

$\Box$

Now let us analyze
\begin{align*}
\int\limits_{\omega_A} \zeta^{n-1}\Big (\Big(P_l-\dfrac{1}{t_*W}\Delta_l-\dfrac{i\pi\xi}{n}\hat\nu_l-\zeta+
O((l-1)^4/W^{2})\Big)^{-1}f_{0}, \bar f_{0}\Big) dz
\end{align*}
\begin{itemize}
\item {\bf localized regime: $\bf W\ll n.$} In this regime $l=1$, so we need to study
\begin{align*}
&\int\limits_{\omega_A} \zeta^{n-1}\Big (\Big(P_1-\dfrac{i\pi\xi}{n}\hat\nu_1-\zeta\Big)^{-1}f_{0}, \bar f_{0}\Big) d\zeta\\
&=-2\pi i\cdot \|f_0\|^2
\Big(\Big(P_1-\dfrac{i\pi\xi}{n}\hat\nu_1\Big)^{n-1} 1,1\Big).
\end{align*}
But since $\phi_0=1$ and $\hat\nu_1=P_1\hat \nu P_1$, $\hat\nu\cdot 1=\phi_1$ (see (\ref{nu}) and Proposition \ref{p:K(U)}), we obtain
\[
(P_1-\dfrac{i\pi\xi}{n}\hat\nu_1\Big) 1=1-\dfrac{i\pi\xi}{n} P_1\phi_1=1,
\]
which implies 
\[
\Big(\Big(P_1-\dfrac{i\pi\xi}{n}\hat\nu_1\Big)^{n-1} 1,1\Big)=1,
\]
thus Theorem \ref{thm:1} in the regime $W\ll n$.


\item {\bf critical regime: $\bf n=C_*W.$} 
Again we need to study $\Big( K_{0}^{n-1} f_0,f_0\Big)$ with 
\begin{align}\label{K_0}
 K_0=P_l-\dfrac{1}{t_*W}\Delta_l-\dfrac{i\pi\xi}{n}\hat\nu_l+O(l^4/W^2)=P_l-\dfrac{C^*}{n}\Delta_l-\dfrac{i\pi\xi}{n}\hat\nu_l+O(\log^4n/n^2),
\end{align}
where $C^*=C_*/t_*$. 
It is enough  to prove
\begin{lemma}\label{l:lim}
Given (\ref{K_*xi}), if $n=C_*W$, $l=[\log W]$ we have
\[
( K_0^{n-1}1,1)\to (e^{-C^*\Delta-i\xi\pi\hat\nu}1,1),\quad n,W\to\infty,
\]
with $\hat\nu$, $\Delta$ as in Theorem \ref{thm:1}.
\end{lemma}
Similar Lemma is proved in \cite{TSh:ChP_crit}, but for the sake of completeness we repeat the proof here.

\noindent{\it Proof of Lemma \ref{l:lim}.}  Notice that 
\begin{align*}
K_{0}=P_l-n^{-1}C^*\Delta_l-\dfrac{i\xi\pi}{n} \hat\nu_l+O(\log^4n/n^{2})=P_l e^{-n^{-1}(C^*\Delta_l+i\xi\pi \hat\nu_l)+O(\log^4n/n^{2})}P_l.
\end{align*}
 Thus
\begin{align*}
K_0^{n-1}=P_le^{-C^*\Delta_l-i\xi\pi \hat\nu_l}P_l+O(\log^4n/n),
\end{align*}
ans so
\[
(K_0^{n-1}1,1)=(e^{-C^*\Delta_l-i\xi\pi \hat\nu_l}1,1)+O(\log^4n/n).
\]
Consider the basis $\{\phi_{j}\}$ of (\ref{phi_j}). In this basis
Laplace operator $\Delta$ is diagonal, and operator $\hat\nu$ is  three diagonal (since it corresponds to the multiplication by $2x^2-1$, see
(\ref{nu}) and (\ref{3_diag})).
To simplify notations, let $F$ be an operator of multiplication by $(i \pi\xi\nu)$ and $\tilde\Delta=C^*\Delta$. Set
\begin{align*}
&D=\tilde\Delta+F,\\ \notag
&D^{(l)}=\tilde\Delta+F^{(l)},
\end{align*}
where $F^{(l)}$ be the matrix $F$ where we put $F_{l-1,l}=F_{l,l-1}=0$. It is evident that (recall $\phi_0=1$)
\[
(e^{-D^{(l)}}\phi_0,\phi_0)=\Big(e^{-P_lD P_l}\phi_0,\phi_0\Big)=(e^{-P_l(C^*\Delta+i\xi\pi \nu)P_l}1,1).
\]
Thus we are left to prove that
\begin{equation}\label{D-D_l}
\Big(\big(e^{-D}-e^{-D^{(l)}}\big)\phi_0,\phi_0\Big)\to 0.
\end{equation}
Notice that both $e^{-D}$, $e^{-D^{(l)}}$ are bounded operators, and $|F|\le C$, $|F^{(l)}|\le C$ . We will use the well-known Duhamel formula
\begin{equation}\label{Duh}
e^{-tA_1}-e^{-tA_2}=\int\limits_{0}^te^{-(t-s)A_2}(A_1-A_2)e^{-sA_1}ds.
\end{equation}
For $A_1=D$, $A_2=D^{(l)}$ and $t=1$ it gives
\begin{align*}
&\Big|\big(e^{-D}-e^{-D^{(l)}}\big)\phi_0\Big|=\Big|\int_0^1e^{-(1-s)D^{(l)}}(F-F^{(l)})e^{-s D}\phi_0\,ds\Big|\\
&=\Big|\int_0^1e^{-(1-s)D^{(l)}}(F_{l-1} \cdot E_{l-1,l}+F_{l}\cdot E_{l,l-1})e^{-s D}\phi_0\,ds\Big|\\
&=\Big|\int_0^1e^{-(1-s)D^{(l)}}\Big(F_{l}\phi_{l}\big(e^{-s D}\phi_0,\phi_{l-1}\big)+F_{l-1}\phi_{l-1}\big(e^{-s D}\phi_0,\phi_{l}\big)\Big)\,ds\Big|\\
&\le C\Big(\big|\big(e^{-s D}\phi_0,\phi_{l-1}\big)\big|+\big|\big(e^{-s D}\phi_0,\phi_{l}\big)\big|\Big).
\end{align*}
Here $E_{l-1,l}$ is an operator whose matrix in the basis $\{\phi_{j}\}$ has 1 at $(l-1,l)$ place and zeros everywhere else,
and $E_{l,l-1}$ is defined in a similar way. $F_{l-1}$, $F_{l}$ are $(l-1,l)$ and $(l,l-1)$ elements of the matrix $F$ in the same basis.

Now let us bound $\big|\big(e^{-s D}\phi_0,\phi_l\big)\big|$. To this end apply Duhamel's formula (\ref{Duh}) $p=[l/2]$ times with
$A_1=D$ and $A_2=\tilde \Delta$. We obtain
\begin{align*}
&\big(e^{-s D}\phi_0,\phi_l\big)=\sum\limits_{j=1}^p \int_{s_1+..+s_j\le s}\big(e^{-s_1\Delta}Fe^{-s_2\Delta}F\ldots e^{-s_j\Delta}\phi_0,\phi_l)\, ds_1..ds_j\\
&+\int_{s_1+..+s_p\le s}\big(e^{-s_1D}Fe^{-s_2\Delta}F\ldots e^{-s_p\Delta}\phi_0,\phi_l)\, ds_1..ds_p.
\end{align*}
Since $e^{-s\Delta}$ is diagonal in the basis $\{\phi_{j}\}$, and $F$ is only three diagonal, the expression $e^{-s_1\Delta}
Fe^{-s_2\Delta}F\ldots e^{-s_j\Delta}\phi_0$ is in the linear span of $\{\phi_{k}\}_{k=0}^j$, and thus the sum above is $0$. 
Hence
\begin{align*}
&\Big|\big(e^{-s D}\phi_0,\phi_l\big)\Big|\le \Big|\int_{s_1+..+s_p\le s}\big(e^{-s_1D}Fe^{-s_2\Delta}F\ldots e^{-s_p\Delta}\phi_0,\phi_l)\, ds_1..ds_p\Big|\\
&\le
C^l\Big|\int_{s_1+..+s_p\le s} ds_1..ds_p\Big|=\dfrac{C^ls^l}{l!}\le C_1e^{-l\log l}\to 0,
\end{align*}
which finishes the proof of (\ref{D-D_l}).

$\Box$

\item {\bf delocalized regime: $\bf W\gg n.$} Since in this regime $l^4/W^{2}=C\log^4n/n^2$, we get
\begin{align*}
&\int\limits_{\omega_A} \zeta^{n-1}\Big (\Big(P_l-\dfrac{C_*}{W}\Delta_l-\dfrac{i\pi\xi}{n}\hat\nu_l-\zeta+
O(l^4/W^{2})\Big)^{-1}f_{0}, \bar f_{0}\Big) d\zeta\\
&=\int\limits_{\omega_A} \zeta^{n-1}\Big (\Big(P_l-\dfrac{C_*}{W}\Delta_l-\dfrac{i\pi\xi}{n}\hat\nu_l-\zeta\Big)^{-1}f_{0}, \bar f_{0}\Big) d\zeta+O(\log^4n/n),
\end{align*}
Hence we need to study
\begin{align*}
|f_0|^2\int\limits_{\omega_A} z^{n-1}\Big (\Big(P_l-\dfrac{C_*}{W}\Delta_l-\dfrac{i\pi\xi}{n}\hat\nu_l-z\Big)^{-1}1, 1\Big) dz
\end{align*}
Now let us define
\begin{equation*}
m=\sqrt[3]{\frac{W}{n}}, \quad m\to\infty,
\end{equation*}
in order to get
\begin{equation*}
\dfrac{m^2n}{W}=\dfrac{1}{m}\to 0.
\end{equation*}
Set
\begin{align*}\notag
&G(z)=(P_l-\dfrac{C_*}{W}\Delta_l-\dfrac{i\pi\xi}{n}\hat\nu_l-z)^{-1},\quad G^{(m)}(z)=(P_m-\dfrac{C_*}{W}\Delta_m-\dfrac{i\pi\xi}{n}\hat\nu_{m}-z)^{-1},\\
&G^{(m,l)}(z)=(P_l-\dfrac{C_*}{W}\Delta_l-\dfrac{i\pi\xi}{n}\hat\nu_{l,m}-z)^{-1},
\end{align*}
where $\hat\nu_{l,m}$ has the same matrix as a tridiagonal operator $\hat\nu_l$ but with $(m,m+1)$ and $(m+1,m)$ elements equal to $0$,
and $P_m$ is a projection on $\{\phi_j\}_{j\le m}$, $\Delta_m=P_m\Delta P_m$, $\hat\nu_m=P_m \hat\nu P_m$.  Notice that $\hat\nu_{l,m}$
has a block diagonal structure with blocks $m\times m$ and $(l-m)\times (l-m)$, thus
\begin{equation}\label{G_m}
( G^{(m,l)}(z)1,1)=(G^{(m)}(z)1,1).
\end{equation}
We are going to prove
\begin{equation}\label{obr_m}
\int\limits_{\omega_A} z^{n-1}\Big (G(z)\cdot 1, 1\Big) dz=\int\limits_{\omega_A} z^{n-1}\Big (G^{(m,l)}(z)\cdot 1, 1\Big) dz+O(1/m).
\end{equation}
Then, if we define
\begin{equation}\label{G_0}
G_0^{(m)}(z)=(P_m-\dfrac{i\pi\xi}{n}\hat\nu_{m}-z)^{-1},
\end{equation} 
we can write using (\ref{G_m}) and the standard resolvent identity
\begin{align*}
\Big (G^{(m,l)}(z)\cdot 1, 1\Big) =(G^{(m)}(z)1,1)=(G^{(m)}_0(z)1,1)+(G^{(m)}(z)\big(\dfrac{C_*}{W}\Delta_m\big)G^{(m)}_0(z)1,1).
\end{align*}
But
\[
\|\dfrac{C_*}{W}\Delta_m\|\le \frac{Cm^2}W\le\frac{1}{mn}, 
\]
hence, since both resolvent can be bounded by $|z-1|^{-1}$, we get
\begin{align}\label{bound}
\Big|\int\limits_{\omega_A} z^{n-1} (G^{(m)}(z)\big(\dfrac{C_*}{W}\Delta_m\big)G^{(m)}_0(z)1,1) dz\Big|\le \dfrac{C}{mn}\int\limits_{\omega_A}\dfrac{|dz|}{|z-1|^2}\le \frac{C}{m},
\end{align}
where we have used
\begin{equation}\label{int_z-1^2}
\int\limits_{\omega_A}\dfrac{|dz|}{|z-1|^2}\le Cn.
\end{equation}
Now (\ref{G_m}) -- (\ref{bound}) imply
\begin{align*}
&|f_0|^2\int\limits_{\omega_A} z^{n-1} (G(z)1, 1) dz=|f_0|^2\int\limits_{\omega_A} z^{n-1}(G^{(m)}_0(z)1,1) dz+O(1/m)\\
&=-2\pi i\cdot |f_0|^2\cdot 
\Big(\Big(P_m-\dfrac{i\pi\xi}{n}\hat\nu_m\Big)^{n-1} 1,1\Big)+O(1/m).
\end{align*}
Since $\hat\nu$ is bounded,  we can easily change $\hat\nu_m$ to $\hat\nu$ and use
\[
\Big(1-\dfrac{i\pi\xi}{n}\hat\nu\Big)^{n-1} =e^{-i\pi\xi\hat\nu}+O(1/n),
\]
which implies Theorem \ref{thm:1}.

Therefore we are left to prove (\ref{obr_m}). First we will need a  bound 
\begin{lemma}\label{l:res_el} For $|z|\ge 1+A/n$ we have
\[
|G_{ij}(z)|\le \dfrac{C}{|z-1|} e^{-\delta |i-j|},
\]
where $C$ and $\delta$ depends only on $A$. Same is true for $G^{(m,l)}(z)$.
\end{lemma}
Notice that since $\hat\nu$ is bounded 3-diagonal matrix (see (\ref{3_diag})), and $P_l-\dfrac{C_*}{W}\Delta_l$ is diagonal,
Lemma \ref{l:res_el} follows from the standard Combes-Thomas arguments (see, e.g., \cite{Pa-Sh}, Ch 13, Proposition 13.13.1)).

%
Using the resolvent identity we can write
\begin{align}\notag
\int\limits_{\omega_A} z^{n-1}\Big (G(z)\cdot 1, 1\Big) dz=&\int\limits_{\omega_A} z^{n-1}\Big (G^{(m,l)}(z)\cdot 1, 1\Big) dz\\
&+
\int\limits_{\omega_A} z^{n-1}\Big (G(z)\Big(\frac{\delta \hat\nu}{n}\Big)G^{(m,l)}(z)\cdot 1, 1\Big) dz,\label{res_id2}
\end{align}
where $\delta \hat\nu=i\pi\xi(\hat\nu_l-\hat\nu_{l,m})$, i.e. the matrix with only two non-zero elements $(m,m+1)$ and $(m+1,m)$. 
Rewrite
\begin{multline}\label{res_el}
\Big (G(z)\Big(\frac{\delta \hat\nu}{n}\Big)G^{(m,l)}(z) 1, 1\Big) \\=\frac{(\delta \hat\nu)_{m,m+1}}{n}G_{m+1,0}^{(m,l)}(z)G_{0m}^*(z)+\frac{(\delta \hat\nu)_{m+1,m}}{n}
G_{m,0}^{(m,l)}(z) G_{0,m+1}^*(z).
\end{multline}
But according to Lemma \ref{l:res_el}
\[
|G_{m+1,0}^{(m,l)}(z)|\le \dfrac{C}{|z-1|} e^{-\delta m},
\]
and similar bounds hold for other resolvent elements in (\ref{res_el}). Thus
\begin{align*}
\Big|\int\limits_{\omega_A}z^{n-1}\Big (G(z)\Big(\frac{\delta \hat\nu}{n}\Big)G^{(m,l)}(z) 1, 1\Big) dz\Big|
\le \dfrac{Ce^{-2\delta m}}{n}\int\limits_{\omega_A}\dfrac{|dz|}{|z-1|^2}\le Ce^{-2\delta m},
\end{align*}
where we have used (\ref{int_z-1^2}). This and (\ref{res_id2}) yield (\ref{obr_m}).
\end{itemize}

\section{Analysis of $I_+$ and $I_-$}
Since the integrals $I_+$ and $I_-$, we can consider $I_+$ only.
In this case we will consider $\{F_i\}$ of Proposition \ref{p:int_repr} like 
$\mathring{Sp}(2)$ matrix which is in $W^{-1/2}$-neigbourhood  of $a_+I_4$. Then  $F_i$ can be  parametrized as $F_i=a_+(I+i\theta_+ X_i/\sqrt{W})$, where
$X_i$ is a quaternion Hermitian matrix 
\[
X_j=\left(
\begin{array}{llll}
\tilde a_{1j}&\tilde w_{j1}&0&\tilde w_{j2}\\
\overline{\tilde{w}}_{j1}&\tilde a_{j2}&-\tilde{w}_{j2}&0\\
0&-\overline{\tilde{w}}_{j2}&\tilde a_{j1}&\overline{\tilde{w}}_{j1}\\
\overline{\tilde{w}}_{j2}&0&\tilde w_{j1}&\tilde a_{j2}
\end{array}
\right),\quad 
\]
where $\tilde w_{j1}=(x_j+iy_j)/\sqrt{2}$, $\tilde w_{j2}=(p_j+iq_j)/\sqrt{2}$.
This change transforms the measure $dF_i$  to
\[
\dfrac{(ia_+\theta_+)^6}{4}W^{-3}d\tilde  a_{1i} d\tilde  a_{2i} d\tilde  x_i \, d\tilde  y_i\,d\tilde  p_i d\tilde  q_i.
\]

We need to keep the same $\tilde C_{n,W}' $ as in (\ref{F_rep1}), so in the parametrization above the operator $K_+^+$ has the form
\begin{equation*}
K_+^+(X,X') 
= \beta^2 A_{a}^+(\bar a,x,y,p,q; \bar a',x',y',p',q') (1+o(1)),
\end{equation*}
where
\begin{align*}
&A_{a}^+(\bar a,x,y,p,q; \bar a',x',y',p',q') \\
&\qquad\qquad =A^+(\widetilde a_1,\widetilde a_1')A^+(\widetilde a_2,\widetilde a_2')
A^+(x,x')A^+(y,y')A^+(p,p')A^+(q,q').
\end{align*}
with $A^+$ of (\ref{K_a*}).
Similarly to Lemma \ref{l:A_ab} one can get that  the largest eigenvalue of  $A_{a}^+$ is $\beta^2(\lambda_0^+)^6+O(W^{-1})$ (see (\ref{lam_0})), and the next eigenvalue is smaller then $\beta^2(\lambda_0^+)^6(1-\delta)$.
Remember that we have normalization $\lambda_0(K_{\pm})^{-1}$, and $\lambda_0(K_{\pm})=\lambda_0^+\lambda_0^-+O(1/W)$ (see Lemma \ref{l:A_ab}).
But according to (\ref{lam_bound}) $\beta |\lambda_0^+|^2<1$, thus
\[
\|\lambda_0(K_{\pm})^{-1} K_+\|<1-\delta,
\]
and so
\[
I_+=O(e^{-cn}).
\]

\section{Appendix A: SUSY techniques}
Here we provide the basic formulas and definitions of SUSY approach used in Section 2.

Let us consider two sets of formal variables
$\{\psi_j\}_{j=1}^n,\{\overline{\psi}_j\}_{j=1}^n$, which satisfy the anticommutation
conditions
\begin{equation}\label{anticom}
\psi_j\psi_k+\psi_k\psi_j=\overline{\psi}_j\psi_k+\psi_k\overline{\psi}_j=\overline{\psi}_j\overline{\psi}_k+
\overline{\psi}_k\overline{\psi}_j=0,\quad j,k=1,\ldots,n.
\end{equation}
Note that this definition implies $\psi_j^2=\overline{\psi}_j^2=0$.
These two sets of variables $\{\psi_j\}_{j=1}^n$ and $\{\overline{\psi}_j\}_{j=1}^n$ generate the Grassmann
algebra $\mathfrak{A}$. Taking into account that $\psi_j^2=0$, we have that all elements of $\mathfrak{A}$
are polynomials of $\{\psi_j\}_{j=1}^n$ and $\{\overline{\psi}_j\}_{j=1}^n$ of degree at most one
in each variable. We can also define functions of
the Grassmann variables. Let $\chi$ be an element of $\mathfrak{A}$, i.e.
\begin{equation}\label{chi}
\chi=a+\sum\limits_{j=1}^n (a_j\psi_j+ b_j\overline{\psi}_j)+\sum\limits_{j\ne k}
(a_{j,k}\psi_j\psi_k+
b_{j,k}\psi_j\overline{\psi}_k+
c_{j,k}\overline{\psi}_j\overline{\psi}_k)+\ldots.
\end{equation}
For any
sufficiently smooth function $f$ we define by $f(\chi)$ the element of $\mathfrak{A}$ obtained by substituting $\chi-a$
in the Taylor series of $f$ at the point $a$. Since $\chi$ is a polynomial of $\{\psi_j\}_{j=1}^n$,
$\{\overline{\psi}_j\}_{j=1}^n$ of the form (\ref{chi}), according to (\ref{anticom}) there exists such
$l$ that $(\chi-a)^l=0$, and hence the series terminates after a finite number of terms and so $f(\chi)\in \mathfrak{A}$.

For example, we have
\begin{align}\notag
 &\exp\{a\,\overline{\psi}_1\psi_1\}=1+a\,\overline{\psi}_1\psi_1+(a\,\overline{\psi}_1\psi_1)^2/2+\ldots
 =1+a\,\overline{\psi}_1\psi_1,\\ \notag
&\exp\{a_{11}\overline{\psi}_1\psi_1+a_{12}\overline{\psi}_1\psi_2+
a_{21}\overline{\psi}_2\psi_1+a_{22}\overline{\psi}_2\psi_2\}=1+ a_{11}\overline{\psi}_1\psi_1\\
\label{ex_12} &+a_{12}\overline{\psi}_1\psi_2+ a_{21}\overline{\psi}_2\psi_1+a_{22}\overline{\psi}_2\psi_2+
(a_{11}\overline{\psi}_1\psi_1+a_{12}\overline{\psi}_1\psi_2\\ \notag &+
a_{21}\overline{\psi}_2\psi_1+a_{22}\overline{\psi}_2\psi_2)^2/2+\ldots=1+
a_{11}\overline{\psi}_1\psi_1+a_{12}\overline{\psi}_1\psi_2+ a_{21}\overline{\psi}_2\psi_1\\ \notag
&+a_{22}\overline{\psi}_2\psi_2+(a_{11}a_{22}-a_{12}a_{21})\overline{\psi}_1\psi_1\overline{\psi}_2\psi_2.
\end{align}
Following Berezin \cite{Ber}, we define the operation of
integration with respect to the anticommuting variables in a formal
way:
\begin{equation}\label{int_gr}
\intd d\,\psi_j=\intd d\,\overline{\psi}_j=0,\quad \intd
\psi_jd\,\psi_j=\intd \overline{\psi}_jd\,\overline{\psi}_j=1,
\end{equation}
and then extend the definition to the general element of $\mathfrak{A}$ by
the linearity. A multiple integral is defined to be a repeated
integral. Assume also that the ``differentials'' $d\,\psi_j$ and
$d\,\overline{\psi}_k$ anticommute with each other and with the
variables $\psi_j$ and $\overline{\psi}_k$. Thus, according to the definition, if
$$
f(\psi_1,\ldots,\psi_k)=p_0+\sum\limits_{j_1=1}^k
p_{j_1}\psi_{j_1}+\sum\limits_{j_1<j_2}p_{j_1,j_2}\psi_{j_1}\psi_{j_2}+
\ldots+p_{1,2,\ldots,k}\psi_1\ldots\psi_k,
$$
then
\begin{equation}\label{int}
\intd f(\psi_1,\ldots,\psi_k)d\,\psi_k\ldots d\,\psi_1=p_{1,2,\ldots,k}.
\end{equation}

   Let $A$ be an ordinary Hermitian matrix with positive real part. The following Gaussian
integral is well-known
\begin{equation}\label{G_C}
\intd \exp\Big\{-\sum\limits_{j,k=1}^nA_{jk}z_j\overline{z}_k\Big\} \prod\limits_{j=1}^n\dfrac{d\,\Re
z_jd\,\Im z_j}{\pi}=\dfrac{1}{\mdet A}.
\end{equation}
One of the important formulas of the Grassmann variables theory is the analog of this formula for the
Grassmann algebra (see \cite{Ber}):
\begin{equation}\label{G_Gr}
\int \exp\Big\{-\sum\limits_{j,k=1}^nA_{jk}\overline{\psi}_j\psi_k\Big\}
\prod\limits_{j=1}^nd\,\overline{\psi}_jd\,\psi_j=\mdet A,
\end{equation}
where $A$ now is any $n\times n$ matrix.

For $n=1$ and $2$ this formula  follows immediately from (\ref{ex_12}) and (\ref{int}).

We will also need the following proposition
\begin{proposition}({\bf see \cite{SupB:08} and references therein})\label{p:supboz}\\
Let $\psi_j=(\psi_{j1},\ldots,\psi_{jm})^t$, $j=1,\ldots, p$ be the Grassman vectors, and let $F$ be some function that depends only on combinations
\begin{align*}
\psi^+\psi&:=\Big\{\sum\limits_{\alpha=1}^m \bar{\psi}_{j\alpha}\psi_{k\alpha}\Big\}_{j,k=1}^p,\quad
\psi\psi^t&:=\Big\{\sum\limits_{\alpha=1}^m\psi_{j\alpha}\psi_{k\alpha}\Big\}_{j,k=1}^p,\quad
\psi^+\bar\psi&:=\Big\{\sum\limits_{\alpha=1}^m\bar \psi_{j\alpha}\bar \psi_{k\alpha}\Big\}_{j,k=1}^p
\end{align*}
and set
\[
d\Psi=\prod\limits_{j=1}^p\prod\limits_{\alpha=1}^m d\bar{\psi}_{j\alpha} d\psi_{j\alpha}.
\]
Assume also that $m\ge p$. Then
\begin{equation*}
\int F\left(\begin{array}{cc}
\psi^+\psi & \psi^+\bar \psi\\
\psi^t\psi & \psi^+\psi
\end{array}
\right)d\Phi d\Psi=C_{p,m}\int F(Q)\cdot \mdet^{-m/2} Q \,d\mu(Q),
\end{equation*}
where $C_{p,m}$ is some constant depending on $p$ and $m$, $Q\in Sp(p)$, and $d\mu(Q)$ is a Haar measure over $Sp(p)$.
\end{proposition}

\section{Appendix B: Proof of Proposition \ref{p:K(U)}}

The first part of Proposition \ref{p:K(U)} follows from the standard representation theory arguments and can be found e.g.
in \cite{Helg}, Ch.5. The recurrent relation (\ref{3_diag}) follows from the recurrent relation for hypergeometric functions, see
e.g. \cite{Ab-St:65}.

Notice also that operator $\hat\nu$ correspond to the multiplication on $c(2x^2-1)$ with $x=\sqrt{S(Q)}$ (see (\ref{nu,s(Q)})).
Thus (\ref{3_diag}) gives that $\hat\nu\phi_0$ is proportional to $\phi_1$, which implies (\ref{nu1,1}).

To get the asymptotic expression (\ref{l_j}) for the eigenvalues of $K_*$ we need the Itzykson-Zuber formula of the integration over $\mathring{Sp}(2)$
(for the proof see, e.g., \cite{TSh:ChP_sym})
\begin{proposition}\label{p:Its-Z}
If $p\ne 0$, then
\begin{equation}\label{Its-Zub_s1}
\int_{\mathring{Sp}(2)} \exp\{-p\,S\big(Q(Q')^*\big)\} d\,\mu(Q')
=\dfrac{6}{p^2}\,
\Big(1-2/p+
e^{-p}\big(1+2/p\big)\Big).
\end{equation}
Moreover,
\begin{equation}\label{Int_del}
\int_{\mathring{Sp}(2)} \exp\{i\pi\xi-2i\pi\xi S\big(Q\big)\} d\,\mu(Q)
=DS(\pi\xi).
\end{equation}
\end{proposition}
Given (\ref{3_diag}), it is easy to check that $P_{2n}(0)=(-1)^n$,
and the coefficient at $x^2$ of $P_{2n}$ is $(-1)^{n-1}n(n+3)/2$. Therefore
\begin{align*}
&\lambda_j(t)=\dfrac{p^2}{6}\int_{\mathring{Sp}(2)} \exp\{-p\,S\big(Q\big)\} (-1)^jP_{2j}(Q)\,d\mu(Q)\\
&=\dfrac{p^2}{6}\int_{\mathring{Sp}(2)} \exp\{-p\,S\big(Q\big)\}\Big(1-\dfrac{j(j+3)}{2}S(Q)+\ldots\Big)\,d\mu(Q)\\
&=1-\dfrac{2}{p}+\dfrac{p^2}{6}\Big(-\dfrac{12}{p^3}\Big(1-\dfrac{2}{p}\big)+\dfrac{6}{p^2}\cdot \dfrac{2}{p^2} \Big)\cdot \dfrac{j(j+3)}{2}+O((j^2/Wt)^2)\\
&=1-\dfrac{(j+1)(j+2)}{Wt}+O((j^2/Wt)^2)
\end{align*}
with $p=Wt$. Here we used $j(j+3)+2=(j+1)(j+2)$, (\ref{Its-Zub_s1}), and
\[
\int_{\mathring{Sp}(2)} \exp\{-p\,S\big(Q\big)\} (S(Q))^k\,d\mu(Q)=(-1)^k\Big(\dfrac{d}{dp}\Big)^k\int_{\mathring{Sp}(2)} \exp\{-p\,S\big(Q\big)\}\,d\mu(Q).
\]

\end{document}